\documentclass[conference]{IEEEtran}
\IEEEoverridecommandlockouts
\usepackage{cite}
\usepackage{amsmath,amssymb,amsfonts}
\usepackage{algorithmic}
\usepackage{graphicx}
\usepackage{textcomp}
\usepackage{xcolor}
\usepackage{setspace}
\usepackage{graphicx}
\usepackage{amsthm}
\usepackage{lscape}
\usepackage{latexsym}
\usepackage{bm}
\usepackage{epstopdf}
\usepackage{bbm}
\usepackage{cite}
\usepackage{url}
\usepackage{color}
\usepackage[caption=false,font=footnotesize]{subfig}
\usepackage{graphicx}
\usepackage{lmodern}
\usepackage[euler]{textgreek}
\usepackage{multirow}

\usepackage{pgfplots}
\usepackage{pgfplotstable}
\pgfplotsset{compat=1.17}
\newcounter{marknumber}
\pgfplotsset{
    error bars/every nth mark/.style={
        /pgfplots/error bars/draw error bar/.prefix code={
            \pgfmathtruncatemacro\marknumbercheck{mod(floor(\themarknumber/2),#1)}
            \ifnum\marknumbercheck=0
            \else
                \begin{scope}[opacity=0]
            \fi
        },
/pgfplots/error bars/draw error bar/.append code={
            \ifnum\marknumbercheck=0
            \else
                \end{scope}
            \fi
            \stepcounter{marknumber}    
        }
    }
}
\def\BibTeX{{\rm B\kern-.05em{\sc i\kern-.025em b}\kern-.08em
    T\kern-.1667em\lower.7ex\hbox{E}\kern-.125emX}}
    
\begin{document}

\newtheorem{Lemma}{Lemma}
\newtheorem{Definition}{Definition}

\title{Deep Joint Source-Channel and Encryption Coding: Secure Semantic Communications}
\author{Tze-Yang Tung, Deniz G\"und\"uz\\
Department of Electrical and Electronics Engineering, Imperial College London\\
\{tze-yang.tung14, d.gunduz\}@imperial.ac.uk
}

\maketitle

\begin{abstract}
Deep learning driven joint source-channel coding (JSCC) for wireless image or video transmission, also called \emph{DeepJSCC}, has been a topic of interest recently with very promising results.
The idea is to map similar source samples to nearby points in the channel input space such that, despite the noise introduced by the channel, the input can be recovered with minimal distortion.
In DeepJSCC, this is achieved by an autoencoder architecture with a non-trainable channel layer between the encoder and decoder.
DeepJSCC has many favorable properties, such as better end-to-end distortion performance than its separate source and channel coding counterpart as well as graceful degradation with respect to channel quality.
However, due to the inherent correlation between the source sample and channel input, DeepJSCC is vulnerable to eavesdropping attacks.
In this paper, we propose the first DeepJSCC scheme for wireless image transmission that is secure against eavesdroppers, called \emph{DeepJSCEC}.
\emph{DeepJSCEC} not only preserves the favorable properties of DeepJSCC, it also provides security against chosen-plaintext attacks from the eavesdropper, without the need to make assumptions about the eavesdropper's channel condition, or its intended use of the intercepted signal.
Numerical results show that \emph{DeepJSCEC} achieves similar or better image quality than separate source coding using BPG compression, AES encryption, and LDPC codes for channel coding, while preserving the graceful degradation of image quality with respect to channel quality.
We also show that the proposed encryption method is problem agnostic, meaning it can be applied to other end-to-end JSCC problems, such as remote classification, without modification.
Given the importance of security in modern wireless communication systems, we believe this work brings DeepJSCC schemes much closer to adoption in practice.
\end{abstract}

\section{Introduction}
\label{sec:intro}

Shannon's separation theorem \cite{Shannon:1948} is the foundation of most modern communication systems.
It states that source coding, which reduces redundancies in the source signal, can be implemented separately from channel coding, which reintroduces redundancies in order to protect against channel distortions, without loss of optimality if given enough tolerance for end-to-end delay.
In the context of wireless image transmission, commonly used compression schemes, such as JPEG2000 \cite{christopoulos_jpeg2000_2000} and BPG \cite{Bellard:BPG}, allow for the reduction in communication load with minimal loss in reconstruction quality, before a channel code is applied to ensure reliable transmission of the compressed bits to the destination. 
This modular architecture also makes incorporating encryption into the design very easy, as the compressed bits can simply be encrypted using a known encryption scheme, such as the Advanced Encryption Standard (AES) \cite{heronAdvancedEncryptionStandard2009}.

However, in practical applications we are limited to finite blocklengths, and it is known that combining the two coding steps, that is, joint source-channel coding (JSCC), can achieve lower distortion for a given finite blocklength than separate source and channel coding \cite{Gastpar:IT:03, kostina_lossy_2013, gallager_information_1968}. 
Despite being a challenging problem, recently it was shown in \cite{Eirina:TCCN:19} that deep neural networks (DNNs) can be used to break the complexity barrier of designing JSCC schemes for wireless image transmission.
The scheme, called \emph{DeepJSCC}, showed appealing properties, such as lower end-to-end distortion for a given channel blocklength compared to state-of-the-art digital compression schemes \cite{Kurka:IZS2020},
flexibility to adapt to different source or channel models \cite{Eirina:TCCN:19,Kurka:IZS2020}, 
ability to exploit channel feedback \cite{Kurka:deepjsccf:jsait}, 
capability to produce adaptive-bandwidth transmission schemes \cite{Kurka:BandwidthAgile:TWComm2021},
and adaptivity to channel input constellation constraints \cite{tungDeepJSCCQConstellationConstrained2022a}.
An extension to the video domain was also recently shown in \cite{tungDeepWiVeDeepLearningAidedWireless2021}.
Importantly, graceful degradation of end-to-end distortion with respect to decreasing channel quality means that DeepJSCC is able to avoid the \textit{cliff-effect} that all separation-based schemes suffer from,
which refers to the phenomenon where the image becomes un-decodable when the channel quality falls below a certain threshold resulting in unreliable transmission.

Despite the success of DeepJSCC schemes, one major problem that has not been addressed is the security.
DeepJSCC schemes learn the encoding and decoding functions from scratch by directly mapping the source signal to the modulated channel input, without conversion to bits at all.
This means that the channel input symbols are directly correlated with the input sample.
It is this property that enables graceful degradation of end-to-end distortion, but it also means known encryption schemes cannot be applied directly.
Therefore, an eavesdropper intercepting the communication channel can theoretically also recover the input sample unchallenged.
Although encryption based on permutation or scrambling has been proposed in the past for analog communications \cite{mosaChaoticEncryptionSpeech2011, borujeniSpeechEncryptionBased2000}, these methods do not provide sufficient security for modern use cases.

In this paper, we propose a DeepJSCC scheme for wireless image transmission that is secure against eavesdroppers, called \emph{deep joint source-channel and encryption coding (DeepJSCEC)}.
We leverage the affine property of the learning with errors (LWE) problem, first introduced in \cite{regevLatticesLearningErrors}, which has been the foundation of many proposed post-quantum cryptographic systems \cite{lindnerBetterKeySizes2011}, to construct a secure wireless image transmission scheme around it.
We emphasize that the security of cryptographic systems based on the LWE problem is still a subject of debate, although there are very strong arguments that suggest its security, which we will outline in Sec. \ref{subsec:encryption}. 
This paper does not attempt to prove the security of such schemes.
Instead, we will assume their security and build a DeepJSCC scheme exploiting them. 
The contributions of this paper are summarized as follows:
\begin{enumerate}
    \item We propose the first DNN designed JSCC scheme for wireless image transmission that is secure against eavesdroppers, called \emph{DeepJSCEC}.
    \item \emph{DeepJSCEC} not only retains the properties that prior works on DeepJSCC have shown, such as graceful degradation of image quality against varying channel quality and lower end-to-end distortion, it is also secure against chosen-plaintext attacks from the eavesdropper.
    \item The cryptographic scheme used in \emph{DeepJSCEC} is a public-key encryption scheme, meaning anyone can send encrypted messages to the intended user using the public key, without the eavesdropper obtaining the message.
    Moreover, new keys can be generated independent of the encoder and decoder models, meaning there is no need to retrain them.
    This makes the scheme very practical for real world use.
    \item Unlike other works that focus on wiretap channel codes
    \cite{liuSecureNestedCodes2007, klincLDPCCodesGaussian2011, fritschekReinforceSecurityModelFree2021, fritschekDeepLearningBased2020, fritschekDeepLearningGaussian2019, besserWiretapCodeDesign2020}, 
    we do not require any assumptions about the eavesdropper's channel quality, or its intended use of the intercepted signal.
    \item Numerical results show that \emph{DeepJSCEC} achieves similar or better image quality than separation-based schemes employing BPG \cite{Bellard:BPG} for source coding, LDPC \cite{gallager_low-density_1962} codes for channel coding and AES \cite{heronAdvancedEncryptionStandard2009} for encryption.
    \item We also show that the proposed encryption method is problem agnostic, meaning it can be applied to other end-to-end JSCC problems, such as remote classification, without modification.
\end{enumerate}

\section{Related Works}
\label{sec:related_works}

Secrecy and privacy in data communication and sharing has been studied extensively in the literature \cite{shannonCommunicationTheorySecrecy1949, wynerWiretapChannel1975, blochPhysicalLayerSecurityInformation2011, dupincalmonPrivacyStatisticalInference2012, zamaniDesignFrameworkStrongly2021, makhdoumiInformationBottleneckPrivacy2014, rassouliPerfectPrivacy2021, rassouliOptimalUtilityPrivacyTradeOff2020, erdemirPrivacyAwareCommunicationWiretap2022, erdemirActivePrivacyUtilityTradeOff2021, marchioroAdversarialNetworksSecure2020}.
These works typically take on an information theoretic formulation, where the objective is to ensure the intended communication link between two users is able to achieve its objective (e.g., reconstruct an image), while an eavesdropper cannot infer a certain type of information about the source (e.g., color, gender,...etc.).

A separate line of work focuses on the design of channel codes, called wiretap channel codes, that exploit the physical characteristics of the legitimate receiver's channel over the eavesdropper's in order to allow communication with secrecy guarantees \cite{liuSecureNestedCodes2007, klincLDPCCodesGaussian2011, lingSemanticallySecureLattice2014, wongSecretSharingLDPCCodes2011}.
Recently, DNNs have also been utilized to learn wiretap channel codes using an autoencoder architecture \cite{besserWiretapCodeDesign2020, fritschekDeepLearningBased2020, fritschekDeepLearningGaussian2019, fritschekReinforceSecurityModelFree2021}.
However, many of these works either require very restrictive channel constellations \cite{wongSecretSharingLDPCCodes2011}, are non-constructive \cite{lingSemanticallySecureLattice2014}, or require the eavesdropper's channel to be significantly worse than the legitimate receiver's \cite{liuSecureNestedCodes2007, klincLDPCCodesGaussian2011, fritschekReinforceSecurityModelFree2021, fritschekDeepLearningBased2020, fritschekDeepLearningGaussian2019, besserWiretapCodeDesign2020}.
Moreover, as they follow the separation of source and channel coding design, these codes are suboptimal in general for finite blocklengths, and suffer from the cliff-effect as well.
Information theoretic security and privacy over a wiretap channel was investigated in \cite{erdemirPrivacyAwareCommunicationWiretap2022} and \cite{marchioroAdversarialNetworksSecure2020}. 
However, just like the wiretap channel codes, the adversary's channel quality is assumed to be significantly worse.

The problem of joint source-channel coding and secrecy has been previously investigated in \cite{magliJointSourceChannel2007} and \cite{sinaieLowComplexityJoint2010}.
However, these schemes are based primarily on randomization and permutation, making them susceptible to chosen-plaintext attacks.
Specifically, in \cite{magliJointSourceChannel2007}, a scheme based on arithmetic coding \cite{mackayInformationTheoryInference2003} with forbidden symbols and randomly permuted intervals was introduced. 
However, due to the simplicity of the encryption, it can be broken by comparing output pairs corresponding to inputs that differ in exactly one symbol.
Similarly, in \cite{sinaieLowComplexityJoint2010}, random lengths of forbidden symbols and randomly placed dummy symbols were used to provide security but is not secure against chosen-plaintext attacks.

In this paper, we will consider DeepJSCC schemes for wireless image transmission that are secure under a specific definition of security, called \emph{ciphertext indistinguishability under chosen-plaintext attack} (IND-CPA), where instead of ensuring that no information about the source can be extracted from the channel codeword by an eavesdropper, we will ensure that the eavesdropper cannot extract the source information \emph{within probabilistic polynomial-time}.
Importantly, our scheme does not require any assumptions about the eavesdropper's channel.
To the best of our knowledge, this is the first work that solves the joint source-channel coding and secrecy problem for DNN driven JSCC of images, which offers not only security guarantees but also superior image quality over separation-based schemes and graceful degradation of end-to-end performance with respect to channel condition.

\section{Problem Statement}
\label{sec:problem_def}

\begin{figure}
    \centering
    \includegraphics[width=\linewidth]{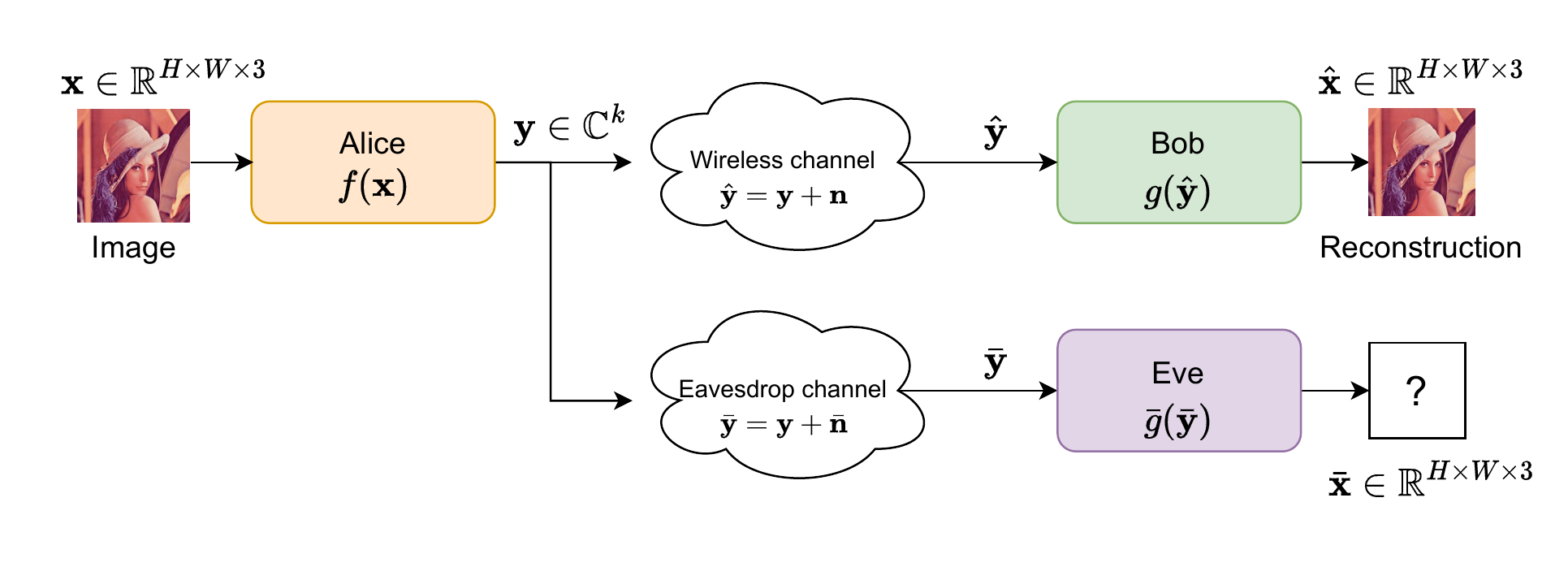}
    \caption{Diagram illustrating problem definition.}
    \label{fig:problem_def}
\end{figure}

We consider the problem of wireless image transmission over an additive white Gaussian noise (AWGN) channel, where Alice wants to send an image to Bob without Eve, the eavesdropper, obtaining a good estimate of the image based on a chosen metric.
Formally, Alice sends an image $\mathbf{x} \in \{0,...,255\}^{H\times W\times C}$ (where $H$, $W$ and $C$ represent the image's height, width and color channels, respectively) using an encoder function $f:\{0,...,255\}^{H\times W\times C} \mapsto \mathbb{C}^k$.
For an RGB image, $C = 3$ for each of the color channels.
Letting $\mathbf{y} = f(\mathbf{x})$, we impose an average transmit power constraint $\bar{P}$, such that
\begin{equation}
    \frac{1}{k}\sum_{i=1}^k |y_i|^2 \leq \bar{P},
\end{equation}
where $y_i$ is the $i$th element of vector $\mathbf{y}$.
The channel input vector $\mathbf{y}$ is transmitted through a noisy channel, with the transfer function $\hat{\mathbf{y}} = \Upsilon(\mathbf{y}) =  \mathbf{y} + \mathbf{n}$, where $\mathbf{n} \sim CN(0,\sigma^2\mathbf{I}_{k\times k})$ is a complex Gaussian vector with dimensionality $k$. 
Bob receives the channel output and decodes it with the function $g:\mathbb{C}^k \mapsto \{0,...,255\}^{H\times W\times C}$ to produce a reconstruction of the input $\hat{\mathbf{x}}=g(\hat{\mathbf{y}})$.
Eve, an eavesdropper, has access to the transmitted message $\mathbf{y}$ via an eavesdropping channel $\bar{\mathbf{y}} = \bar{\Upsilon}(\mathbf{y}) =  \mathbf{y} + \bar{\mathbf{n}}$, where $\bar{\mathbf{n}} \sim CN(0,\sigma_e^2\mathbf{I}_{k\times k})$ is the channel noise observed by Eve with variance $\sigma_e^2$. 
We will also assume that Eve has access to known pairs of $(\mathbf{x}, \bar{\mathbf{y}})$ such that she can try to find a decoder $\bar{\mathbf{x}} = \bar{g}(\bar{\mathbf{y}})$.
This is called a \emph{chosen-plaintext} attack.
A diagram illustrating the problem is shown in Fig. \ref{fig:problem_def}.

Given the above problem definition, we define the channel SNR between Alice and Bob as
\begin{equation}
    \text{SNR}_b = 10\log_{10} \left( \frac{\bar{P}}{{\sigma}^2} \right) \text{ dB},
\end{equation}
and the corresponding channel SNR for Eve as
\begin{equation}
    \text{SNR}_e = 10\log_{10} \left( \frac{\bar{P}}{{\sigma}_e^2} \right) \text{ dB}.
\end{equation}
We also define the \emph{bandwidth compression ratio} as
\begin{equation}
    \rho = \frac{k}{H\times W\times C}~\text{channel symbols/pixel},
\end{equation}
where a smaller number reflects more compression.

We will consider three distortion metrics for images. 
The first is the mean squared error (MSE),
\begin{equation}
    \text{MSE}(\mathbf{x},\hat{\mathbf{x}})=||\mathbf{x}-\hat{\mathbf{x}}||_2^2.
    \label{eq:mse_loss}
\end{equation}
A common complimentary metric to the MSE distortion to measure image reconstruction quality is the peak signal-to-noise ratio (PSNR) defined as
\begin{equation}
    \text{PSNR}(\mathbf{x},\hat{\mathbf{x}})=\log_{10}\bigg(\frac{A^2}{\text{MSE}(\mathbf{x},\hat{\mathbf{x}})}\bigg)~\text{dB},
    \label{eq:psnr_def}
\end{equation}
where $A$ is the maximum possible value for a given pixel. 
For a 24 bit RGB pixel, $A=255$.
Maximizing this metric corresponds to minimizing the MSE.

The second is the structural similarity index measure (SSIM), defined as
\begin{equation}
    \text{SSIM}(\mathbf{x},\hat{\mathbf{x}})=
    \left(\frac{2\mu_{\mathbf{x}}\mu_{\hat{\mathbf{x}}}+v_1}{\mu_{\mathbf{x}}^2+\mu_{\hat{\mathbf{x}}}^2+v_1}\right) 
    \left(\frac{2\sigma_{\mathbf{x}}\sigma_{\hat{\mathbf{x}}}+v_2}{\sigma_{\mathbf{x}}^2+\sigma_{\hat{\mathbf{x}}}^2+v_2}\right),
    \label{eq:ssim_loss}
\end{equation}
where $\mu_{\mathbf{x}}$, $\sigma^2_{\mathbf{x}}$, $\sigma^2_{\mathbf{x}\hat{\mathbf{x}}}$ are the mean and variance of $\mathbf{x}$, and the covariance between $\mathbf{x}$ and $\hat{\mathbf{x}}$, respectively, and $v_1$, $v_2$ are coefficients for numeric stability.

The last metric is the multi-scale structural similarity index (MS-SSIM) metric, defined as
\begin{equation}
\begin{aligned}
    &\text{MS-SSIM}(\mathbf{x},\hat{\mathbf{x}}) \\
    &= [l_M(\mathbf{x},\hat{\mathbf{x}})]^{\alpha_M}\prod_{j=1}^M[a_j(\mathbf{x},\hat{\mathbf{x}})]^{\beta_j}[b_j(\mathbf{x},\hat{\mathbf{x}})]^{\gamma_j},
\end{aligned}
    \label{eq:msssim_loss}
\end{equation}
where
\begin{align}
    l_M(\mathbf{x},\hat{\mathbf{x}}) &= \frac{2\mu_{\mathbf{x}}\mu_{\hat{\mathbf{x}}}+v_1}{\mu_{\mathbf{x}}^2+\mu_{\hat{\mathbf{x}}}^2+v_1},\\
    a_j(\mathbf{x},\hat{\mathbf{x}}) &= \frac{2\sigma_{\mathbf{x}}\sigma_{\hat{\mathbf{x}}}+v_2}{\sigma_{\mathbf{x}}^2+\sigma_{\hat{\mathbf{x}}}^2+v_2},\\
    b_j(\mathbf{x},\hat{\mathbf{x}}) &= \frac{\sigma_{\mathbf{x}\hat{\mathbf{x}}}+v_3}{\sigma_{\mathbf{x}}\sigma_{\hat{\mathbf{x}}}+v_3},
\end{align}
$\mu_{\mathbf{x}}$, $\sigma^2_{\mathbf{x}}$, $\sigma^2_{\mathbf{x}\hat{\mathbf{x}}}$ are the mean and variance of $\mathbf{x}$, and the covariance between $\mathbf{x}$ and $\hat{\mathbf{x}}$, respectively.
$v_1$, $v_2$, and $v_3$ are coefficients for numeric stability;
$\alpha_M$, $\beta_j$, and $\gamma_j$ are the weights for each of the components.
Each $a_j(\cdot,\cdot)$ and $b_j(\cdot,\cdot)$ is computed at a different downsampled scale of $\mathbf{x}$ and $\hat{\mathbf{x}}$.
We use the default parameter values of ($\alpha_M$, $\beta_j$, $\gamma_j$) provided by the original paper \cite{wang_multiscale_2003}.
MS-SSIM has been shown to perform better in approximating the human visual perception than the more simplistic structural similarity index (SSIM) on different subjective image and video databases.
However, due to the need for downsampling, MS-SSIM will only be used for images larger than $64 \times 64$ during training and testing.
Both the SSIM and the MS-SSIM have a maximum value of 1, which means maximizing either metric corresponds to minimizing $1 - \text{(MS-)SSIM}$.

For the security metric, we will use \emph{ciphertext indistinguishability under chosen-plaintext attack} (IND-CPA) to evaluate the security of the scheme.
It can be defined formally in the form of a public-key cryptographic game as follows
\begin{Definition}
Let $\mathcal{K}$ be a key generation function, $\mathcal{E}$ be an encryption function, and $\mathcal{D}$ be a decryption function that defines a cryptographic scheme
$\mathcal{S}\mathcal{E} = (\mathcal{K}, \mathcal{E}, \mathcal{D})$.
Let an adversary (Eve), be computationally bounded by a probabilistic polynomial-time Turing machine.
The game for testing IND-CPA is defined as:
\begin{enumerate}
    \item Initialize $(P_k, S_k) \stackrel{\$}{\leftarrow} \mathcal{K}$, $b \stackrel{\$}{\leftarrow} Ber(0.5)$, where $Ber(0.5)$ is a Bernoulli distribution with success probability $0.5$ and $\stackrel{\$}{\leftarrow}$ refers to the sampling operation.
    Here, $P_k$ is the public key and $S_k$ is the secret key.
    Eve does not know $S_k$ and $b$, but knows $P_k$.
    \item Knowing the public key $P_k$, Eve is able to generate any number of ciphertexts within polynomial-time.
    \item Eve chooses two plaintext messages $M_0$ and $M_1$ of equal length, and provides them and $P_k$ to an oracle, who computes the ciphertext $C \stackrel{\$}{\leftarrow} \mathcal{E}(M_b, P_k)$, returning the ciphertext of one of the chosen plaintext messages determined by $b$.
    \item Based on $C$, Eve tries to determine $M_{b^\prime}$ as the plaintext used to compute $C$.
\end{enumerate}
\label{def:ind_cpa}
\end{Definition}
The advantage that Eve has for a given encryption scheme $\mathcal{S}\mathcal{E}$ is defined as
\begin{equation}
    \text{Adv}_{\mathcal{S}\mathcal{E}}^{\text{IND-CPA}} = 2 P(b^\prime = b) - 1.
\end{equation}
As such, an advantage of 0 means $P(b^\prime = b) = 0.5$, which corresponds to a random guess by Eve.
In the context of our problem, we wish to allow Eve to at best learn a decoding function $\bar{g}$ to reconstruct $\bar{g}(\bar{\mathbf{y}}) \sim P(\mathbf{x})$, where $P(\mathbf{x})$ is the prior distribution on the source images, within probabilistic polynomial-time.
The goal is then to design an encoder, decoder pair $(f, g)$, such that the reconstruction quality $d(\mathbf{x}, \hat{\mathbf{x}})$, measured by either Eqn. (\ref{eq:psnr_def}), (\ref{eq:ssim_loss}), or (\ref{eq:msssim_loss}) is maximized while giving Eve an advantage $\text{Adv}_{\mathcal{S}\mathcal{E}}^{\text{IND-CPA}} \approx 0$ based on IND-CPA. 
We note that this definition is similar to Shannon's notion of perfect secrecy but rather than assuming the adversary can have infinite computing capability, we consider adversaries restricted by probabilistic polynomial-time Turing machines.

\section{Proposed Solution}
\label{sec:proposed_solution}

An input image $\mathbf{x}$ is mapped with a non-linear encoder function $f_{\boldsymbol{\theta}}:\mathbb{R}^{H\times W\times C}\mapsto \mathbb{R}^k$, parameterized by $\boldsymbol{\theta}$, into a latent vector $\mathbf{z}=f_{\boldsymbol{\theta}}(\mathbf{x})$. 
Each value in the latent vector $\mathbf{z}$ is then quantized into $N$ uniform quantization levels, with centroids $\mathcal{C}_q=\{q_1,...,q_N\}$, via the quantizer $q_{\mathcal{C}_q}:\mathbb{R}^k \mapsto \mathcal{C}_q^k$, which we will define in Sec. \ref{subsec:quantization}. 
That is, $\bar{\mathbf{z}}=q_{\mathcal{C}_q}(\mathbf{z})$, with $\bar z_i \in \mathcal{C}_q$, where $\bar z_i$ is the $i$th element of the quantized vector $\bar{\mathbf{z}}$.
We will refer to $\bar{\mathbf{z}}$ as the plaintext.

The plaintext $\bar{\mathbf{z}}$ is then encrypted using a public-key encryption scheme $E:\mathbb{Z}_p^k \times \mathbb{Z}_p^{n_1 \times k} \mapsto \mathbb{Z}_{p}^{k}$, producing a ciphertext $\mathbf{c} = E(\bar{\mathbf{z}}, \mathbf{P}(\mathbf{S}))$, where $\mathbf{P}(\mathbf{S}) \in \mathbb{Z}_p^{n_1 \times k}$ is the public key and a function of the secret key $\mathbf{S} \in \mathbb{Z}_p^{n_2 \times k}$. 
The constants $(n_1, n_2)$ will be determined later.
Let there also be a corresponding decryption scheme $D(\mathbf{c}, \mathbf{P}(\mathbf{S}), \mathbf{S}) = \mathbf{z}^\prime$.
We will assume for now that the secret key $\mathbf{S}$ cannot be inferred from the public key $\mathbf{P}(\mathbf{S})$.
The entire encryption procedure will be described in Sec. \ref{subsec:encryption}.

The ciphertext $\mathbf{c}$ is then modulated using a constellation $\mathcal{C} = \{c_1, \dots, c_p\}$ of order $p$, by mapping each $c_j \in \mathbb{Z}_p$ to the corresponding constellation point in $\mathcal{C}$, producing channel input $\mathbf{y} \in \mathcal{C}^k$.
To ensure that the power constraint is met, we choose the constellation points such that the average power of the constellation points assuming uniform probability is
\begin{equation}
    \bar{P} = \frac{1}{p}\sum_{j=1}^p |c_j|^2.
\end{equation}
The channel input $\mathbf{y}$ is then transmitted through an AWGN channel, producing the channel output $\hat{\mathbf{y}} = \mathbf{y} + \mathbf{n}$. 

At the receiver, the likelihood of each received value is first computed as
\begin{equation}
    l_{i}^j = P(\hat{y}_i | c_j),
\label{eq:likelihood}
\end{equation}
where $\hat{y}_i$ is the $i$th element in $\hat{\mathbf{y}}$ and $c_j$ is the $j$th constellation point in $\mathcal{C}$.
We then compute the noisy ciphertext $\hat{\mathbf{c}} \in \mathbb{R}^{k}$ by computing the softmax weighted sum of values in $\mathbb{Z}_p$ based on the likelihood vector $\mathbf{l}_i = [ l_i^1, \dots, l_i^p]$ for $i=1,...,k$ as
\begin{equation}
    \hat{c}_i = \sum_{j=1}^{p} \frac{e^{\sigma_l l_i^j}}{\sum_{n=1}^p e^{\sigma_l l_i^n}} (j-1),
\label{eq:noisy_ct}
\end{equation}
where $\hat{c}_i$ is the $i$th element in $\hat{\mathbf{c}}$ and $\sigma_l$ is the parameter controlling the weighting of the likelihoods.
Let us denote the channel noise in the ciphertext space as $\mathbf{n}_c = \hat{\mathbf{c}} - \mathbf{c}$.
Given the noisy ciphertext $\hat{\mathbf{c}}$, we can obtain a noisy plaintext by using the corresponding decryptor $D: \mathbb{R}^{k} \mapsto \mathbb{R}_p^{k}$ and the secret key $\mathbf{S}$ as $\mathbf{z}^\prime = D(\hat{\mathbf{c}}, \mathbf{P}(\mathbf{S}), \mathbf{S})$.

The resultant noisy plaintext $\mathbf{z}^\prime$ is generally too noisy to be decoded effectively.
Therefore, we compute the softmax weighted sum of the symbols in $\mathcal{C}_q$ based on their $l_2$ distances from $z^\prime_i \in \mathbf{z}^\prime$; 
that is,
\begin{equation}
    \hat{z_i} = \sum_{j=1}^N \frac{e^{-d^\prime_{ij}}}{\sum_{n=1}^N e^{- d^\prime_{in}}} q_j,
\end{equation}
where $d^\prime_{ij} = ||z^\prime_i-q_j||^2_2$, is the $l_2$ distance between $z^\prime_i$ and the centroid $q_j \in \mathcal{C}_q$.

Finally, the estimate $\hat{\mathbf{z}}$ is passed through a non-linear decoder function $g_{\boldsymbol{\phi}}:\mathbb{R}^k\mapsto\mathbb{R}^{H\times W\times C}$, parameterized by $\boldsymbol{\phi}$, to produce a reconstruction of the input $\hat{\mathbf{x}}=g_{\boldsymbol{\phi}}(\hat{\mathbf{z}})$.

\subsection{Quantization}
\label{subsec:quantization}

Given the encoder output $\mathbf{z}$, we first obtain a ``hard" quantization, which simply maps element $z_i\in\mathbf{z}$ to the nearest symbol in $\mathcal{C}_q$.
This forms the quantized latent vector $\bar{\mathbf{z}}$.
However, this operation is not differentiable.
In order to obtain a differentiable approximation of the hard quantization operation, we will use the  ``soft" quantization approach, proposed in \cite{Agustsson:softQuant:NIPS2017}. 
In this approach, each quantized value is generated as the softmax weighted sum of the symbols in $\mathcal{C}$ based on their $l_2$ distances from $z_i$; 
that is,
\begin{equation}
\label{eq:soft_quantization}
    \tilde{z_i} = \sum_{j=1}^N \frac{e^{-\sigma_q d_{ij}}}{\sum_{n=1}^N e^{-\sigma_q d_{in}}} q_j,
\end{equation}
where $\sigma_q$ is a parameter controlling the ``hardness'' of the assignment, and $d_{ij} = ||z_i-q_j||^2_2$, is $l_2$ distance between the latent value $z_i$ and the centroid $q_j$.
As such, in the forward pass, the quantizer uses the hard quantization $\bar{\mathbf{z}}$, and in the backward pass, the gradient from the soft quantization $\tilde{\mathbf{z}}$ is used to update $\boldsymbol{\theta}$.
That is,
\begin{equation}
    \frac{\partial\bar{\mathbf{z}}}{\partial \mathbf{z}} =
    \frac{\partial\tilde{\mathbf{z}}}{\partial \mathbf{z}}.
\end{equation}
To ensure that the quantized values lie within $\mathbb{Z}_p$, we define the centroids to be $\mathcal{C}_q = \{\lfloor\frac{ip}{N}\rfloor\}_{i=0}^{N-1}$.

\subsection{Encryption}
\label{subsec:encryption}

The foundation of the encryption scheme is the learning with error (LWE) problem, which was initially introduced in \cite{regevLatticesLearningErrors}.
The problem is parameterized by dimensions $n_1, n_2 \geq 1$ and an integer modulus $p \geq 2$, as well as an error distribution $\chi$ over $\mathbb{Z}$.
In this paper, we will use a discrete Gaussian distribution $\chi = G_{\mathbb{Z}, \sigma_s}$, which is a probability distribution that assigns probability proportional to $\exp(-\pi x^2 / \sigma_s^2)$ for $x \in \mathbb{Z}$.
The LWE problem stipulates that given a lattice $\mathbf{A} \in \mathbb{Z}_p^{n_1 \times n_2}$ sampled uniformly from $\mathbb{Z}_p$, a secret $\mathbf{s} \in \mathbb{Z}^{n_1}$ and error term $\mathbf{e} \in \mathbb{Z}^{n_2}$ sampled independent and identically distributed (i.i.d.) from $\chi$, it is difficult to recover $\mathbf{s}$ from the pair $(\mathbf{A}, \mathbf{A}^\top \mathbf{s} + \mathbf{e})$ in polynomial-time. 
The difficulty lies in the fact that it is difficult to distinguish pairs of $(\mathbf{A}, \mathbf{A}^\top \mathbf{s} + \mathbf{e})$ from $(\mathbf{A}, \mathbf{b}^\prime)$, where $\mathbf{b}^\prime$ is sampled uniformly from $\mathbb{Z}_p^{n_2}$.
In fact, in \cite{regevLatticesLearningErrors}, it is stated that the best known algorithm for distinguishing $(\mathbf{A}, \mathbf{A}^\top \mathbf{s} + \mathbf{e})$ from $(\mathbf{A}, \mathbf{b}^\prime)$ with an advantage greater than zero requires $2^{O(n_1 / \log n_1)}$ space and time.
This meets the IND-CPA security metric defined in Sec. \ref{sec:problem_def}.

Using the LWE problem, in \cite{lindnerBetterKeySizes2011}, a public key cryptographic system was constructed in the following way.
Let $\mathbf{S} \in \mathbb{Z}_p^{n_2 \times k}$, $\mathbf{U} \in \mathbb{Z}_p^{n_1 \times k}$ be matrices whose values are sampled i.i.d. from a discrete Gaussian distribution $G_{\mathbb{Z}, \sigma_s}$.
Let $\mathbf{P}(\mathbf{S}) = (\mathbf{U} - \mathbf{A}\mathbf{S}, \mathbf{A})$ be the public key and $\mathbf{S}$ as the secret key.
Then, to encrypt with the public key, the ciphertext is computed from a plaintext $\bar{\mathbf{z}} \in \mathbb{Z}_p^k$ as
\begin{align}
    &E(\bar{\mathbf{z}}, \mathbf{P}(\mathbf{S})) \label{eq:encrypt_start} \\
    &= ((\mathbf{U} - \mathbf{A}\mathbf{S})^\top\mathbf{e}_1 + \mathbf{e}_3 + \bar{\mathbf{z}}, ~\mathbf{A}^\top\mathbf{e}_1 + \mathbf{e}_2)~(\text{mod }p) \\
    &\overset{\Delta}{=} (\mathbf{c}, \mathbf{d})~(\text{mod }p), \label{eq:encrypt_end}
\end{align}
where $\mathbf{e} = (\mathbf{e}_1, \mathbf{e}_2, \mathbf{e}_3) \in \mathbb{Z}^{n_1} \times \mathbb{Z}^{n_2} \times \mathbb{Z}^k$ are all sampled i.i.d. from $G_{\mathbb{Z}, \sigma_s}$ as well.
Note that the seed used to generate $\mathbf{S}$ and $\mathbf{U}$ are different from the one used to generate $\mathbf{e}$.
The seed used to generate the error terms $\mathbf{e}$ is public as it is part of the encryption function, whereas the seed used to generate $\mathbf{S}$ and $\mathbf{U}$ is private as it generates the secret key.
In order to decipher, the legitimate recipient, who has the secret key $\mathbf{S}$, computes  
\begin{align}
    \hat{\mathbf{z}} &= D(\mathbf{c}, \mathbf{d}, \mathbf{P}(\mathbf{S}), \mathbf{S}) \\
    &= \mathbf{S}^\top\mathbf{d} + \mathbf{c} ~(\text{mod }p)\\
    &= \mathbf{S}^\top\mathbf{e}_2 + \mathbf{U}^\top\mathbf{e}_1 + \mathbf{e}_3 + \bar{\mathbf{z}} ~(\text{mod }p).
\end{align}
If the error terms $\mathbf{S}^\top\mathbf{e}_2 + \mathbf{U}^\top\mathbf{e}_1 + \mathbf{e}_3$ are sufficiently small, then $\lfloor \hat{\mathbf{z}} \rceil \equiv \bar{\mathbf{z}} ~(\text{mod } p)$.
The conditions for correctness given a set of parameters $(n_1, n_2, \sigma_s)$ can be found in \cite{lindnerBetterKeySizes2011}.

Herein, in order to use this encryption technique in our DeepJSCC system, we will assume that both Alice and Bob have the same seed to generate the error terms in $\mathbf{e}$. 
Therefore, we can assume that $\mathbf{d}$ is available to the receiver prior to transmission since it does not depend on the message $\bar{\mathbf{z}}$.
Given the encryption procedure defined in Eqns. (\ref{eq:encrypt_start}-\ref{eq:encrypt_end}), the receiver receives a noisy ciphertext $\hat{\mathbf{c}} = \mathbf{c} + \mathbf{n}_c$, as outlined above, where $\mathbf{n}_c$ is the noise from the channel in the ciphertext space following Eqns. (\ref{eq:likelihood}-\ref{eq:noisy_ct}).
The decryption is then performed as
\begin{align}
    \mathbf{z}^\prime &= D(\hat{\mathbf{c}}, \mathbf{P}(\mathbf{S}), \mathbf{S}) \\
    &= \mathbf{S}^\top\mathbf{d} + \hat{\mathbf{c}} ~(\text{mod }p)\\
    &= \mathbf{S}^\top\mathbf{e}_2 + \mathbf{U}^\top\mathbf{e}_1 + \mathbf{e}_3 + \bar{\mathbf{z}} + \mathbf{n}_c ~(\text{mod }p).
\end{align}

We note that the effect of the channel noise term $\mathbf{n}_c$ and the decryption error terms $\mathbf{S}^\top\mathbf{e}_2 + \mathbf{U}^\top\mathbf{e}_1 + \mathbf{e}_3$ are additive to the message $\bar{\mathbf{z}}$. 
We can therefore treat them jointly as additive noise and allow the encoder and decoder pair $(f_{\boldsymbol{\theta}}, g_{\boldsymbol{\phi}})$ to learn a JSCC codebook that is optimized for the compound noise from the channel and the encryption scheme. 
We then have a IND-CPA secure DeepJSCC scheme.
Importantly, this scheme does not make any assumptions about Eve's channel quality.
It is secure even if Eve has direct access to the noiseless ciphertext $\mathbf{c}$ and the seed needed to generate the error terms $\mathbf{e}$.
Indeed, we will show via a deep learning-based chosen-plaintext attack in Sec. \ref{sec:results} that Eve is not able to learn a decoding function $\bar{g}$ that can reconstruct the original image $\mathbf{x}$ from the noiseless ciphertext $\mathbf{c}$ with a level of detail significantly better than a random sample from $P(\mathbf{x})$.

\section{Numerical Results}
\label{sec:results}

\begin{table}
\centering
\caption{Summary of parameters}
\label{tab:parameters}
\begin{tabular}{|c|c|c|c|}
\hline
$p$ & $n_1=n_2$  & $\sigma_s$  & $N$  \\ \hline
$4093$ & $192$  & $8.87$ & $16$ \\ \hline
\end{tabular}
\end{table}

\begin{figure*}
    \centering
    \includegraphics[width=0.8\linewidth]{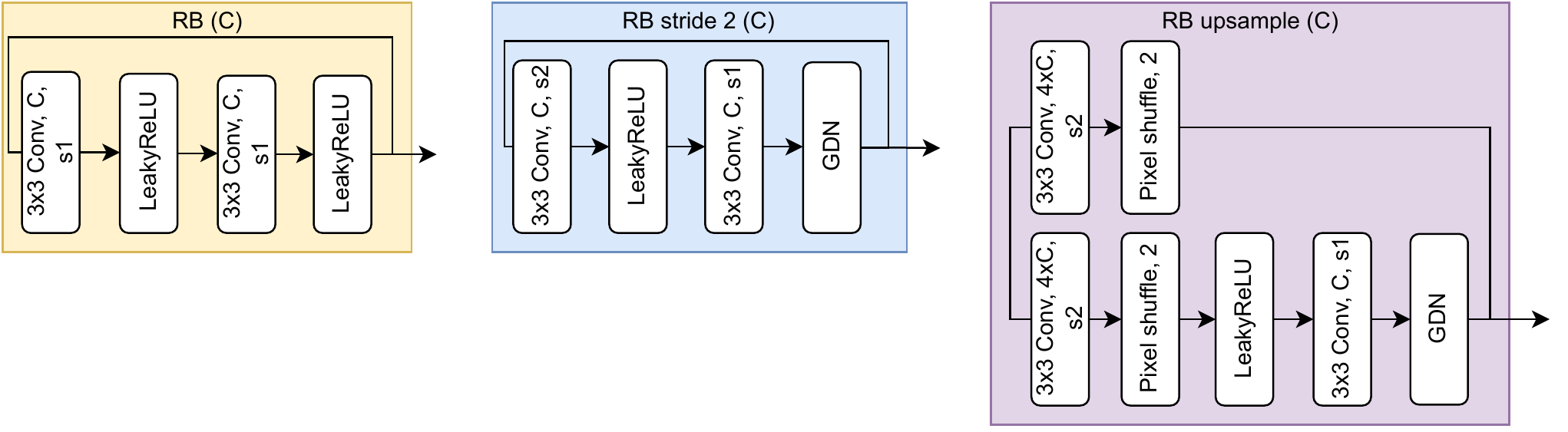}
    \caption{
    Diagram illustrating the residual block types used in the encoder $f_{\boldsymbol{\theta}}$ and decoder $g_{\boldsymbol{\phi}}$.
    }
    \label{fig:residual_blocks}
\end{figure*}

\begin{table*}
\centering
\caption{
Network architectures used for training on CIFAR10 and Tiny ImageNet datasets for the image transmission problem.
The block definitions are shown in Fig. \ref{fig:residual_blocks}.
}
\label{tab:jscc_image_architectures}
\begin{tabular}{|cc|cc|}
\hline
\multicolumn{2}{|c|}{CIFAR10}    & \multicolumn{2}{c|}{Tiny ImageNet}    \\ \hline
\multicolumn{1}{|c|}{$f_{\boldsymbol{\theta}}$} & $g_{\boldsymbol{\phi}}$ & \multicolumn{1}{c|}{$f_{\boldsymbol{\theta}}$} & $g_{\boldsymbol{\phi}}$ \\ \hline
\multicolumn{1}{|c|}{RB stride 2 (256)} & Attention block (256) & \multicolumn{1}{c|}{RB stride 2 (256)} & Attention block (256) \\ \hline
\multicolumn{1}{|c|}{RB (256)} & RB (256) & \multicolumn{1}{c|}{RB (256)} & RB (256) \\ \hline
\multicolumn{1}{|c|}{RB (256)} & RB (256) & \multicolumn{1}{c|}{RB stride 2 (256)} & RB upsample (256) \\ \hline
\multicolumn{1}{|c|}{RB (256)} & RB (256) & \multicolumn{1}{c|}{Attention block (256)} & RB (256) \\ \hline
\multicolumn{1}{|c|}{Attention block (256)} & RB upsample (256) & \multicolumn{1}{c|}{RB (256)} & RB upsample (256) \\ \hline
\multicolumn{1}{|c|}{RB stride 2 (256)} & Attention block (256) & \multicolumn{1}{c|}{RB stride 2 (256)} & Attention block (256) \\ \hline
\multicolumn{1}{|c|}{RB (256)} & RB (256) & \multicolumn{1}{c|}{RB (256)} & RB (256) \\ \hline
\multicolumn{1}{|c|}{RB (256)} & RB (256) & \multicolumn{1}{c|}{RB stride 2 (256)} & RB upsample (256) \\ \hline
\multicolumn{1}{|c|}{RB (256)} & RB (256) & \multicolumn{1}{c|}{RB (256)} & RB (256) \\ \hline
\multicolumn{1}{|c|}{Attention block ($C_{\text{out}}$)} & RB upsample (3) & \multicolumn{1}{c|}{Attention block ($C_{\text{out}}$)} & RB upsample (3) \\ \hline
\end{tabular}
\end{table*}

Herein, we outline the training details used to conduct our experiments.
Firstly, as the encryption method and the modulation/demodulation procedures are not differentiable, the gradient through those operations is skipped during training, such that the gradient of the loss $l(\mathbf{x}, \hat{\mathbf{x}})$ with respect to $\hat{\mathbf{z}}$ is
\begin{equation}
    \frac{\partial l(\mathbf{x}, \hat{\mathbf{x}})}{\partial \hat{\mathbf{z}}} 
    =\frac{\partial l(\mathbf{x}, \hat{\mathbf{x}})}{\partial \bar{\mathbf{z}}}.
\label{eq:grad_skip_training}
\end{equation}
This corresponds to the method described in Sec. \ref{subsec:encryption}, where we jointly treat the compound noise from the encryption scheme and the channel as additive noise.

For the encryption scheme, we use parameters from \cite{lindnerBetterKeySizes2011}, where they computed the values for $p$, $n_1$, $n_2$, and $\sigma_s$ that provide a certain level of security.
The exact values are shown in Table \ref{tab:parameters}.
In \cite{lindnerBetterKeySizes2011}, the authors showed that the parameters in Table \ref{tab:parameters} provide an advantage close to zero ($2^{-32}$) and an eavesdropper would optimistically require in the order of $2^{42}$ seconds to decode the message.
In order to generate the secret key $\mathbf{S}$ and the error terms $(\mathbf{U}, \mathbf{e})$, we use a zero mean Gaussian random variable with variance $\sigma_s^2 / 2\pi$ and round the samples to the nearest integer.
Although increasing the number of quantization levels $N$ would in theory improve the performance, due to the error terms from the cryptographic scheme, increasing $N$ also increases the relative magnitude of the error terms, thus reducing the performance.
As such, we found $N=16$ to be empirically a good trade-off between the resolution and performance.

To aid exploration, the quantization hardness parameter $\sigma_q$ is linearly annealed using the annealing function
\begin{equation}
    \sigma_q^{(i)} = \min \left(200, \sigma_q^{(i-1)} + 5 \left\lfloor \frac{i}{2000} \right\rfloor \right),
\end{equation}
where $i$ is the parameter update step number and we initialize with $\sigma_q^{(0)} = 5$.
As for the soft ciphertext parameter $\sigma_l$, we empirically found that $\sigma_l = 5$ works well in practice.

For the channel input constellation, since the encryption lattice modulus is $4093$, we use a 4096-QAM and simply ignore the last 3 constellation points.
The reason we used this constellation is because it avoids having to design a constellation of a non-standard size, and since the constellation size is sufficiently large, a waste of 3 constellation points is negligible.
Moreover, due to the gradient skipping, as described by Eqn. (\ref{eq:grad_skip_training}), it is not possible to train a custom constellation either.

To compare our scheme with separation-based schemes, we will consider BPG \cite{Bellard:BPG} for source coding and LDPC codes for channel coding.
The LDPC codes we use are from the 802.11ad standard \cite{noauthor_ieee_2012}, with blocklength 672 bits for both the $1/2$ and $3/4$ rate codes.
We use the AES encryption scheme \cite{heronAdvancedEncryptionStandard2009} for the baseline with block and key size of $128$ bits. 
As such, we compress the images to a bit rate such that it is under the rate support by the LDPC code considered and is within an integer multiple of $128$ bits.
Just like the proposed \emph{DeepJSCEC}, we will assume the keys are exchanged prior to transmission.
Clearly, \emph{DeepJSCEC} has a larger overhead than the digital baseline we are considering due to the need to transmit two separate seeds for generating $(\mathbf{S}, \mathbf{U})$ and the error terms $\mathbf{e}$.
However, as will be seen in the subsequent results, this trade-off is worth the gains observed.
\begin{figure} 
\centering
  \subfloat[PSNR
  \label{subfig:cifar10_reconstruction_graceful_psnr}]{%
    \begin{tikzpicture}
        \pgfplotsset{
            legend style={
                font=\fontsize{4}{4}\selectfont,
                at={(1.0,.0)},
                anchor=south east,
            },
            height=0.5\textwidth,
            width=0.5\textwidth,
            xmin=0,
            xmax=17,
            ymin=10,
            ymax=30,
            xtick distance=2,
            ytick distance=2,
            xlabel={$\text{SNR}_b$ (dB)},
            ylabel={PSNR (dB)},
            grid=both,
            grid style={line width=.1pt, draw=gray!10},
            major grid style={line width=.2pt,draw=gray!50},
            every axis/.append style={
                x label style={
                    font=\fontsize{8}{8}\selectfont,
                    at={(axis description cs:0.5,-0.04)},
                    },
                y label style={
                    font=\fontsize{8}{8}\selectfont,
                    at={(axis description cs:-0.08,0.5)},
                    },
                x tick label style={
                    font=\fontsize{8}{8}\selectfont,
                    /pgf/number format/.cd,
                    fixed,
                    fixed zerofill,
                    precision=0,
                    /tikz/.cd
                    },
                y tick label style={
                    font=\fontsize{8}{8}\selectfont,
                    /pgf/number format/.cd,
                    fixed,
                    fixed zerofill,
                    precision=1,
                    /tikz/.cd
                    },
            }
        }
        \begin{axis}
        \addplot[blue, solid, line width=1.5pt, 
        mark=*, mark options={fill=blue, scale=1.1}, 
        error bars/.cd, y dir=both, y explicit, every nth mark=2] 
        table [x=SNR, y=snr_train5, 
        y error=snr_train5_std, col sep=comma]
        {data/djscec_cifar10_bw1667_nembed16_psnr.csv};
        \addlegendentry{\textit{DeepJSCEC} 
        ($\text{SNR}_{\text{Train}}=5dB$)}
        
        \addplot[teal, dashed, line width=1.5pt, 
        mark=triangle*, mark options={fill=teal, scale=1.1, solid}, 
        error bars/.cd, y dir=both, y explicit, every nth mark=2] 
        table [x=SNR, y=snr_train7, 
        y error=snr_train7_std, col sep=comma]
        {data/djscec_cifar10_bw1667_nembed16_psnr.csv};
        \addlegendentry{\textit{DeepJSCEC} 
        ($\text{SNR}_{\text{Train}}=7dB$)}
        
        \addplot[green, solid, line width=1.5pt, 
        mark=*, mark options={fill=green, scale=1.1}, 
        error bars/.cd, y dir=both, y explicit, every nth mark=2] 
        table [x=SNR, y=snr_train10, 
        y error=snr_train10_std, col sep=comma]
        {data/djscec_cifar10_bw1667_nembed16_psnr.csv};
        \addlegendentry{\textit{DeepJSCEC} 
        ($\text{SNR}_{\text{Train}}=10dB$)}
        
        \addplot[brown, dashed, line width=1.5pt, 
        mark=triangle*, mark options={fill=brown, scale=1.1, solid}, 
        error bars/.cd, y dir=both, y explicit, every nth mark=2] 
        table [x=SNR, y=snr_train12, 
        y error=snr_train12_std, col sep=comma]
        {data/djscec_cifar10_bw1667_nembed16_psnr.csv};
        \addlegendentry{\textit{DeepJSCEC} 
        ($\text{SNR}_{\text{Train}}=12dB$)}
        
        \addplot[red, solid, line width=1.5pt, 
        mark=*, mark options={fill=red, scale=1.1}, 
        error bars/.cd, y dir=both, y explicit, every nth mark=2] 
        table [x=SNR, y=snr_train16, 
        y error=snr_train16_std, col sep=comma]
        {data/djscec_cifar10_bw1667_nembed16_psnr.csv};
        \addlegendentry{\textit{DeepJSCEC} 
        ($\text{SNR}_{\text{Train}}=16dB$)}
        
        \addplot[color=black, dashed, line width=1.2pt, mark=*, mark options={fill=black, solid, scale=1.1}, 
        error bars/.cd, y dir=both, y explicit, every nth mark=1] 
        table [x=SNR, y=1_2_bpsk, y error=1_2_bpsk_std, col sep=comma]
        {data/bpg_cifar10_bw1667_psnr.csv};
        \addlegendentry{BPG + LDPC 1/2 BPSK}
        
        \addplot[color=magenta, dashed, line width=1.2pt, 
        mark=*, mark options={fill=magenta, solid, scale=1.1}, 
        error bars/.cd, y dir=both, y explicit, every nth mark=1] 
        table [x=SNR, y=3_4_bpsk, y error=3_4_bpsk_std, col sep=comma]
        {data/bpg_cifar10_bw1667_psnr.csv};
        \addlegendentry{BPG + LDPC 3/4 BPSK}
        
        \addplot[color=black, dashed, line width=1.2pt, 
        mark=triangle*, mark options={fill=black, solid, scale=1.1}, 
        error bars/.cd, y dir=both, y explicit, every nth mark=1] 
        table [x=SNR, y=1_2_qpsk, y error=1_2_qpsk_std, col sep=comma]
        {data/bpg_cifar10_bw1667_psnr.csv};
        \addlegendentry{BPG + LDPC 1/2 QPSK}
        
        \addplot[color=magenta, dashed, line width=1.2pt, 
        mark=triangle*, mark options={fill=magenta, solid, scale=1.1}, 
        error bars/.cd, y dir=both, y explicit, every nth mark=1] 
        table [x=SNR, y=3_4_qpsk, y error=3_4_qpsk_std, col sep=comma]
        {data/bpg_cifar10_bw1667_psnr.csv};
        \addlegendentry{BPG + LDPC 3/4 QPSK}
        
        \addplot[color=black, dashed, line width=1.2pt, 
        mark=square*, mark options={fill=black, solid, scale=1.1}, 
        error bars/.cd, y dir=both, y explicit, every nth mark=1] 
        table [x=SNR, y=1_2_16qam, y error=1_2_16qam_std, col sep=comma]
        {data/bpg_cifar10_bw1667_psnr.csv};
        \addlegendentry{BPG + LDPC 1/2 16QAM}
        \end{axis}
        \end{tikzpicture}
    }
    \\
  \subfloat[SSIM
  \label{subfig:cifar10_reconstruction_graceful_ssim}]{%
    \begin{tikzpicture}
        \pgfplotsset{
            legend style={
                font=\fontsize{4}{4}\selectfont,
                at={(1.0,0.)},
                anchor=south east,
            },
            height=0.5\textwidth,
            width=0.5\textwidth,
            xmin=0,
            xmax=17,
            ymin=0.1,
            ymax=1,
            xtick distance=2,
            ytick distance=0.1,
            xlabel={$\text{SNR}_b$ (dB)},
            ylabel={SSIM},
            grid=both,
            grid style={line width=.1pt, draw=gray!10},
            major grid style={line width=.2pt,draw=gray!50},
            every axis/.append style={
                x label style={
                    font=\fontsize{8}{8}\selectfont,
                    at={(axis description cs:0.5, -0.04)},
                    },
                y label style={
                    font=\fontsize{8}{8}\selectfont,
                    at={(axis description cs:-0.08,0.5)},
                    },
                x tick label style={
                    font=\fontsize{8}{8}\selectfont,
                    /pgf/number format/.cd,
                    fixed,
                    fixed zerofill,
                    precision=0,
                    /tikz/.cd
                    },
                y tick label style={
                    font=\fontsize{8}{8}\selectfont,
                    /pgf/number format/.cd,
                    fixed,
                    fixed zerofill,
                    precision=2,
                    /tikz/.cd
                    },
            }
        }
        \begin{axis}
        \addplot[blue, solid, line width=1.5pt, 
        mark=*, mark options={fill=blue, scale=1.1}, 
        error bars/.cd, y dir=both, y explicit, every nth mark=2] 
        table [x=SNR, y=snr_train5, 
        y error=snr_train5_std, col sep=comma]
        {data/djscec_cifar10_bw1667_nembed16_ssim.csv};
        \addlegendentry{\textit{DeepJSCEC} 
        ($\text{SNR}_{\text{Train}}=5dB$)}
        
        \addplot[teal, dashed, line width=1.5pt, 
        mark=triangle*, mark options={fill=teal, scale=1.1, solid}, 
        error bars/.cd, y dir=both, y explicit, every nth mark=2] 
        table [x=SNR, y=snr_train7, 
        y error=snr_train7_std, col sep=comma]
        {data/djscec_cifar10_bw1667_nembed16_ssim.csv};
        \addlegendentry{\textit{DeepJSCEC} 
        ($\text{SNR}_{\text{Train}}=7dB$)}
        
        \addplot[green, solid, line width=1.5pt, 
        mark=*, mark options={fill=green, scale=1.1}, 
        error bars/.cd, y dir=both, y explicit, every nth mark=2] 
        table [x=SNR, y=snr_train10, 
        y error=snr_train10_std, col sep=comma]
        {data/djscec_cifar10_bw1667_nembed16_ssim.csv};
        \addlegendentry{\textit{DeepJSCEC} 
        ($\text{SNR}_{\text{Train}}=10dB$)}
        
        \addplot[brown, dashed, line width=1.5pt, 
        mark=triangle*, mark options={fill=brown, scale=1.1, solid}, 
        error bars/.cd, y dir=both, y explicit, every nth mark=2] 
        table [x=SNR, y=snr_train12, 
        y error=snr_train12_std, col sep=comma]
        {data/djscec_cifar10_bw1667_nembed16_ssim.csv};
        \addlegendentry{\textit{DeepJSCEC} 
        ($\text{SNR}_{\text{Train}}=12dB$)}
        
        \addplot[red, solid, line width=1.5pt, 
        mark=*, mark options={fill=red, scale=1.1}, 
        error bars/.cd, y dir=both, y explicit, every nth mark=2] 
        table [x=SNR, y=snr_train16, 
        y error=snr_train16_std, col sep=comma]
        {data/djscec_cifar10_bw1667_nembed16_ssim.csv};
        \addlegendentry{\textit{DeepJSCEC} 
        ($\text{SNR}_{\text{Train}}=16dB$)}
        
        \addplot[color=black, dashed, line width=1.2pt, mark=*, mark options={fill=black, solid, scale=1.1}, 
        error bars/.cd, y dir=both, y explicit, every nth mark=1] 
        table [x=SNR, y=1_2_bpsk, y error=1_2_bpsk_std, col sep=comma]
        {data/bpg_cifar10_bw1667_ssim.csv};
        \addlegendentry{BPG + LDPC 1/2 BPSK}
        
        \addplot[color=magenta, dashed, line width=1.2pt, 
        mark=*, mark options={fill=magenta, solid, scale=1.1}, 
        error bars/.cd, y dir=both, y explicit, every nth mark=1] 
        table [x=SNR, y=3_4_bpsk, y error=3_4_bpsk_std, col sep=comma]
        {data/bpg_cifar10_bw1667_ssim.csv};
        \addlegendentry{BPG + LDPC 3/4 BPSK}
        
        \addplot[color=black, dashed, line width=1.2pt, 
        mark=triangle*, mark options={fill=black, solid, scale=1.1}, 
        error bars/.cd, y dir=both, y explicit, every nth mark=1] 
        table [x=SNR, y=1_2_qpsk, y error=1_2_qpsk_std, col sep=comma]
        {data/bpg_cifar10_bw1667_ssim.csv};
        \addlegendentry{BPG + LDPC 1/2 QPSK}
        
        \addplot[color=magenta, dashed, line width=1.2pt, 
        mark=triangle*, mark options={fill=magenta, solid, scale=1.1}, 
        error bars/.cd, y dir=both, y explicit, every nth mark=1] 
        table [x=SNR, y=3_4_qpsk, y error=3_4_qpsk_std, col sep=comma]
        {data/bpg_cifar10_bw1667_ssim.csv};
        \addlegendentry{BPG + LDPC 3/4 QPSK}
        
        \addplot[color=black, dashed, line width=1.2pt, 
        mark=square*, mark options={fill=black, solid, scale=1.1}, 
        error bars/.cd, y dir=both, y explicit, every nth mark=1] 
        table [x=SNR, y=1_2_16qam, y error=1_2_16qam_std, col sep=comma]
        {data/bpg_cifar10_bw1667_ssim.csv};
        \addlegendentry{BPG + LDPC 1/2 16QAM}
        \end{axis}
        \end{tikzpicture}
        }
\caption{
Comparison between \emph{DeepJSCEC} trained on the CIFAR10 dataset for different $\text{SNR}_{\text{Train}}$ and digital baselines using BPG for compression, AES for encryption and LDPC codes for channel coding, for the image transmission problem ($\rho = 1/6$).
}
\label{fig:cifar10_reconstruction_graceful} 
\end{figure}

\begin{figure} 
    \centering
  \subfloat[PSNR
  \label{subfig:tinyimagenet_reconstruction_graceful_psnr}]{%
    \begin{tikzpicture}
        \pgfplotsset{
            legend style={
                font=\fontsize{4}{4}\selectfont,
                at={(1.0,.0)},
                anchor=south east,
            },
            height=0.5\textwidth,
            width=0.5\textwidth,
            xmin=0,
            xmax=17,
            ymin=10,
            ymax=30,
            xtick distance=2,
            ytick distance=2,
            xlabel={$\text{SNR}_b$ (dB)},
            ylabel={PSNR (dB)},
            grid=both,
            grid style={line width=.1pt, draw=gray!10},
            major grid style={line width=.2pt,draw=gray!50},
            every axis/.append style={
                x label style={
                    font=\fontsize{8}{8}\selectfont,
                    at={(axis description cs:0.5,-0.04)},
                    },
                y label style={
                    font=\fontsize{8}{8}\selectfont,
                    at={(axis description cs:-0.08,0.5)},
                    },
                x tick label style={
                    font=\fontsize{8}{8}\selectfont,
                    /pgf/number format/.cd,
                    fixed,
                    fixed zerofill,
                    precision=0,
                    /tikz/.cd
                    },
                y tick label style={
                    font=\fontsize{8}{8}\selectfont,
                    /pgf/number format/.cd,
                    fixed,
                    fixed zerofill,
                    precision=1,
                    /tikz/.cd
                    },
            }
        }
        \begin{axis}
        \addplot[blue, solid, line width=1.5pt, 
        mark=*, mark options={fill=blue, scale=1.1}, 
        error bars/.cd, y dir=both, y explicit, every nth mark=2] 
        table [x=SNR, y=snr_train5, 
        y error=snr_train5_std, col sep=comma]
        {data/djscec_tinyimagenet_bw1667_nembed16_psnr.csv};
        \addlegendentry{\textit{DeepJSCEC} 
        ($\text{SNR}_{\text{Train}}=5dB$)}
        
        \addplot[teal, dashed, line width=1.5pt, 
        mark=triangle*, mark options={fill=teal, scale=1.1, solid}, 
        error bars/.cd, y dir=both, y explicit, every nth mark=2] 
        table [x=SNR, y=snr_train7, 
        y error=snr_train7_std, col sep=comma]
        {data/djscec_tinyimagenet_bw1667_nembed16_psnr.csv};
        \addlegendentry{\textit{DeepJSCEC} 
        ($\text{SNR}_{\text{Train}}=7dB$)}
        
        \addplot[green, solid, line width=1.5pt, 
        mark=*, mark options={fill=green, scale=1.1}, 
        error bars/.cd, y dir=both, y explicit, every nth mark=2] 
        table [x=SNR, y=snr_train10, 
        y error=snr_train10_std, col sep=comma]
        {data/djscec_tinyimagenet_bw1667_nembed16_psnr.csv};
        \addlegendentry{\textit{DeepJSCEC} 
        ($\text{SNR}_{\text{Train}}=10dB$)}
        
        \addplot[brown, dashed, line width=1.5pt, 
        mark=triangle*, mark options={fill=brown, scale=1.1, solid}, 
        error bars/.cd, y dir=both, y explicit, every nth mark=2] 
        table [x=SNR, y=snr_train12, 
        y error=snr_train12_std, col sep=comma]
        {data/djscec_tinyimagenet_bw1667_nembed16_psnr.csv};
        \addlegendentry{\textit{DeepJSCEC} 
        ($\text{SNR}_{\text{Train}}=12dB$)}
        
        \addplot[red, solid, line width=1.5pt, 
        mark=*, mark options={fill=red, scale=1.1}, 
        error bars/.cd, y dir=both, y explicit, every nth mark=2] 
        table [x=SNR, y=snr_train16, 
        y error=snr_train16_std, col sep=comma]
        {data/djscec_tinyimagenet_bw1667_nembed16_psnr.csv};
        \addlegendentry{\textit{DeepJSCEC} 
        ($\text{SNR}_{\text{Train}}=16dB$)}
        
        \addplot[color=black, dashed, line width=1.2pt, 
        mark=*, mark options={fill=black, solid, scale=1.1}, 
        error bars/.cd, y dir=both, y explicit, every nth mark=1] 
        table [x=SNR, y=1_2_bpsk, y error=1_2_bpsk_std, col sep=comma]
        {data/bpg_tinyimagenet_bw1667_psnr.csv};
        \addlegendentry{BPG + LDPC 1/2 BPSK}
        
        \addplot[color=magenta, dashed, line width=1.2pt, 
        mark=*, mark options={fill=magenta, solid, scale=1.1}, 
        error bars/.cd, y dir=both, y explicit, every nth mark=1] 
        table [x=SNR, y=3_4_bpsk, y error=3_4_bpsk_std, col sep=comma]
        {data/bpg_tinyimagenet_bw1667_psnr.csv};
        \addlegendentry{BPG + LDPC 3/4 BPSK}
        
        \addplot[color=black, dashed, line width=1.2pt, 
        mark=triangle*, mark options={fill=black, solid, scale=1.1}, 
        error bars/.cd, y dir=both, y explicit, every nth mark=1] 
        table [x=SNR, y=1_2_qpsk, y error=1_2_qpsk_std, col sep=comma]
        {data/bpg_tinyimagenet_bw1667_psnr.csv};
        \addlegendentry{BPG + LDPC 1/2 QPSK}
        
        \addplot[color=magenta, dashed, line width=1.2pt, 
        mark=triangle*, mark options={fill=magenta, solid, scale=1.1}, 
        error bars/.cd, y dir=both, y explicit, every nth mark=1] 
        table [x=SNR, y=3_4_qpsk, y error=3_4_qpsk_std, col sep=comma]
        {data/bpg_tinyimagenet_bw1667_psnr.csv};
        \addlegendentry{BPG + LDPC 3/4 QPSK}
        
        \addplot[color=black, dashed, line width=1.2pt, 
        mark=square*, mark options={fill=black, solid, scale=1.1}, 
        error bars/.cd, y dir=both, y explicit, every nth mark=1] 
        table [x=SNR, y=1_2_16qam, y error=1_2_16qam_std, col sep=comma]
        {data/bpg_tinyimagenet_bw1667_psnr.csv};
        \addlegendentry{BPG + LDPC 1/2 16QAM}
        \end{axis}
        \end{tikzpicture}
    }
    \\
  \subfloat[MS-SSIM
  \label{subfig:tinyimagenet_reconstruction_graceful_ssim}]{%
    \begin{tikzpicture}
        \pgfplotsset{
            legend style={
                font=\fontsize{4}{4}\selectfont,
                at={(1.0,0.)},
                anchor=south east,
            },
            height=0.5\textwidth,
            width=0.5\textwidth,
            xmin=2,
            xmax=17,
            ymin=0.6,
            ymax=1,
            xtick distance=2,
            ytick distance=0.1,
            xlabel={$\text{SNR}_b$ (dB)},
            ylabel={MS-SSIM},
            grid=both,
            grid style={line width=.1pt, draw=gray!10},
            major grid style={line width=.2pt,draw=gray!50},
            every axis/.append style={
                x label style={
                    font=\fontsize{8}{8}\selectfont,
                    at={(axis description cs:0.5, -0.04)},
                    },
                y label style={
                    font=\fontsize{8}{8}\selectfont,
                    at={(axis description cs:-0.08,0.5)},
                    },
                x tick label style={
                    font=\fontsize{8}{8}\selectfont,
                    /pgf/number format/.cd,
                    fixed,
                    fixed zerofill,
                    precision=0,
                    /tikz/.cd
                    },
                y tick label style={
                    font=\fontsize{8}{8}\selectfont,
                    /pgf/number format/.cd,
                    fixed,
                    fixed zerofill,
                    precision=2,
                    /tikz/.cd
                    },
            }
        }
        \begin{axis}
        \addplot[blue, solid, line width=1.5pt, 
        mark=*, mark options={fill=blue, scale=1.1}, 
        error bars/.cd, y dir=both, y explicit, every nth mark=2] 
        table [x=SNR, y=snr_train5, 
        y error=snr_train5_std, col sep=comma]
        {data/djscec_tinyimagenet_bw1667_nembed16_msssim.csv};
        \addlegendentry{\textit{DeepJSCEC} 
        ($\text{SNR}_{\text{Train}}=5dB$)}
        
        \addplot[teal, dashed, line width=1.5pt, 
        mark=triangle*, mark options={fill=teal, scale=1.1, solid}, 
        error bars/.cd, y dir=both, y explicit, every nth mark=2] 
        table [x=SNR, y=snr_train7, 
        y error=snr_train7_std, col sep=comma]
        {data/djscec_tinyimagenet_bw1667_nembed16_msssim.csv};
        \addlegendentry{\textit{DeepJSCEC} 
        ($\text{SNR}_{\text{Train}}=7dB$)}
        
        \addplot[green, solid, line width=1.5pt, 
        mark=*, mark options={fill=green, scale=1.1}, 
        error bars/.cd, y dir=both, y explicit, every nth mark=2] 
        table [x=SNR, y=snr_train10, 
        y error=snr_train10_std, col sep=comma]
        {data/djscec_tinyimagenet_bw1667_nembed16_msssim.csv};
        \addlegendentry{\textit{DeepJSCEC} 
        ($\text{SNR}_{\text{Train}}=10dB$)}
        
        \addplot[brown, dashed, line width=1.5pt, 
        mark=triangle*, mark options={fill=brown, scale=1.1, solid}, 
        error bars/.cd, y dir=both, y explicit, every nth mark=2] 
        table [x=SNR, y=snr_train12, 
        y error=snr_train12_std, col sep=comma]
        {data/djscec_tinyimagenet_bw1667_nembed16_msssim.csv};
        \addlegendentry{\textit{DeepJSCEC} 
        ($\text{SNR}_{\text{Train}}=12dB$)}
        
        \addplot[red, solid, line width=1.5pt, 
        mark=*, mark options={fill=red, scale=1.1}, 
        error bars/.cd, y dir=both, y explicit, every nth mark=2] 
        table [x=SNR, y=snr_train16, 
        y error=snr_train16_std, col sep=comma]
        {data/djscec_tinyimagenet_bw1667_nembed16_msssim.csv};
        \addlegendentry{\textit{DeepJSCEC} 
        ($\text{SNR}_{\text{Train}}=16dB$)}
        
        \addplot[color=black, dashed, line width=1.2pt, 
        mark=*, mark options={fill=black, solid, scale=1.1}, 
        error bars/.cd, y dir=both, y explicit, every nth mark=1] 
        table [x=SNR, y=1_2_bpsk, y error=1_2_bpsk_std, col sep=comma]
        {data/bpg_tinyimagenet_bw1667_msssim.csv};
        \addlegendentry{BPG + LDPC 1/2 BPSK}
        
        \addplot[color=magenta, dashed, line width=1.2pt, 
        mark=*, mark options={fill=magenta, solid, scale=1.1}, 
        error bars/.cd, y dir=both, y explicit, every nth mark=1] 
        table [x=SNR, y=3_4_bpsk, y error=3_4_bpsk_std, col sep=comma]
        {data/bpg_tinyimagenet_bw1667_msssim.csv};
        \addlegendentry{BPG + LDPC 3/4 BPSK}
        
        \addplot[color=black, dashed, line width=1.2pt, 
        mark=triangle*, mark options={fill=black, solid, scale=1.1}, 
        error bars/.cd, y dir=both, y explicit, every nth mark=1] 
        table [x=SNR, y=1_2_qpsk, y error=1_2_qpsk_std, col sep=comma]
        {data/bpg_tinyimagenet_bw1667_msssim.csv};
        \addlegendentry{BPG + LDPC 1/2 QPSK}
        
        \addplot[color=magenta, dashed, line width=1.2pt, 
        mark=triangle*, mark options={fill=magenta, solid, scale=1.1}, 
        error bars/.cd, y dir=both, y explicit, every nth mark=1] 
        table [x=SNR, y=3_4_qpsk, y error=3_4_qpsk_std, col sep=comma]
        {data/bpg_tinyimagenet_bw1667_msssim.csv};
        \addlegendentry{BPG + LDPC 3/4 QPSK}
        
        \addplot[color=black, dashed, line width=1.2pt, 
        mark=square*, mark options={fill=black, solid, scale=1.1}, 
        error bars/.cd, y dir=both, y explicit, every nth mark=1] 
        table [x=SNR, y=1_2_16qam, y error=1_2_16qam_std, col sep=comma]
        {data/bpg_tinyimagenet_bw1667_msssim.csv};
        \addlegendentry{BPG + LDPC 1/2 16QAM}
        \end{axis}
        \end{tikzpicture}
        }
\caption{
Comparison between \emph{DeepJSCEC} trained on the Tiny ImageNet dataset for different $\text{SNR}_{\text{Train}}$ and digital baselines using BPG for compression, AES for encryption and LDPC codes for channel coding, for the image transmission problem ($\rho = 1/6$).
}
\label{fig:tinyimagenet_reconstruction_graceful} 
\end{figure}

\begin{figure}
    \centering
  \subfloat[PSNR\label{subfig:cifar10_psnr_v_bw}]{%
    \begin{tikzpicture}
        \pgfplotsset{
            legend style={
                font=\fontsize{5}{5}\selectfont,
                at={(1.0,.0)},
                anchor=south east,
            },
            width=0.5\textwidth,
            height=0.3\textwidth,
            xmin=0.05,
            xmax=0.35,
            ymin=8,
            ymax=32,
            xtick distance=0.05,
            ytick distance=4,
            xlabel={$\rho$},
            ylabel={PSNR (dB)},
            grid=both,
            grid style={line width=.1pt, draw=gray!10},
            major grid style={line width=.2pt,draw=gray!50},
            every axis/.append style={
                x label style={
                    font=\fontsize{8}{8}\selectfont,
                    at={(axis description cs:0.5,-0.07)},
                    },
                y label style={
                    font=\fontsize{8}{8}\selectfont,
                    at={(axis description cs:-0.08,0.5)},
                    },
                x tick label style={
                    font=\fontsize{8}{8}\selectfont,
                    /pgf/number format/.cd,
                    fixed,
                    fixed zerofill,
                    precision=2,
                    /tikz/.cd
                    },
                y tick label style={
                    font=\fontsize{8}{8}\selectfont,
                    /pgf/number format/.cd,
                    fixed,
                    fixed zerofill,
                    precision=1,
                    /tikz/.cd
                    },
            }
        }
        \begin{axis}
        \addplot[blue, solid, line width=0.9pt, 
        mark=*, mark options={fill=blue, scale=1.1}, error bars/.cd, y dir=both, y explicit, every nth mark=1] 
        table [x=rho, y=l2, y error=l2_std, col sep=comma]
              {data/djscec_cifar10_nembed16_snr10_rate_dist.csv};
        \addlegendentry{\textit{DeepJSCEC}
        ($\text{SNR}_{\text{Train}}=10dB$)}
        
        \addplot[red, dashed, line width=0.9pt, 
        mark=*, mark options={fill=red, scale=1.1}, error bars/.cd, y dir=both, y explicit, every nth mark=1] 
        table [x=rho, y=l2, y error=l2_std, col sep=comma]
        {data/bpg_cifar10_rate_dist.csv};
        \addlegendentry{BPG LDPC 1/2 + QPSK}
        \end{axis}
        \end{tikzpicture}
    }
    \\
  \subfloat[SSIM\label{subfig:cifar10_ssim_v_bw}]{%
    \begin{tikzpicture}
        \pgfplotsset{
            legend style={
                font=\fontsize{5}{5}\selectfont,
                at={(1.0,0.)},
                anchor=south east,
            },
            width=0.5\textwidth,
            height=0.3\textwidth,
            xmin=0.05,
            xmax=0.35,
            ymin=0.50,
            ymax=1,
            xtick distance=0.05,
            ytick distance=0.1,
            xlabel={$\rho$},
            ylabel={SSIM},
            grid=both,
            grid style={line width=.1pt, draw=gray!10},
            major grid style={line width=.2pt,draw=gray!50},
            every axis/.append style={
                x label style={
                    font=\fontsize{8}{8}\selectfont,
                    at={(axis description cs:0.5,-0.07)},
                    },
                y label style={
                    font=\fontsize{8}{8}\selectfont,
                    at={(axis description cs:-0.08,0.5)},
                    },
                x tick label style={
                    font=\fontsize{8}{8}\selectfont,
                    /pgf/number format/.cd,
                    fixed,
                    fixed zerofill,
                    precision=2,
                    /tikz/.cd
                    },
                y tick label style={
                    font=\fontsize{8}{8}\selectfont,
                    /pgf/number format/.cd,
                    fixed,
                    fixed zerofill,
                    precision=2,
                    /tikz/.cd
                    },
            }
        }
        \begin{axis}
        \addplot[blue, solid, line width=0.9pt, 
        mark=*, mark options={fill=blue, scale=1.1}, error bars/.cd, y dir=both, y explicit, every nth mark=1] 
        table [x=rho, y=ssim, y error=ssim_std, col sep=comma]
        {data/djscec_cifar10_nembed16_snr10_rate_dist.csv};
        \addlegendentry{\textit{DeepJSCEC}
        ($\text{SNR}_{\text{Train}}=10dB$)}
        
        \addplot[red, dashed, line width=0.9pt, 
        mark=*, mark options={fill=red, scale=1.1}, error bars/.cd, y dir=both, y explicit, every nth mark=1] 
        table [x=rho, y=ssim, y error=ssim_std, col sep=comma]
        {data/bpg_cifar10_rate_dist.csv};
        \addlegendentry{BPG LDPC 1/2 + QPSK}
        \end{axis}
        \end{tikzpicture}
        }
  \caption{
  Comparison of \emph{DeepJSCEC} trained on the CIFAR10 dataset to BPG compression for different bandwidth compression ratios $\rho$.
  }
  \label{fig:cifar10_distortion_v_bw} 
\end{figure}
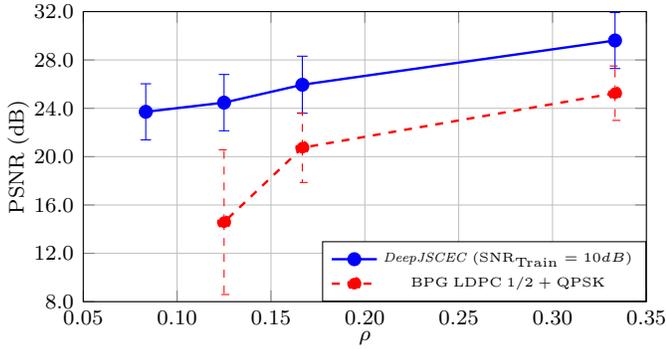
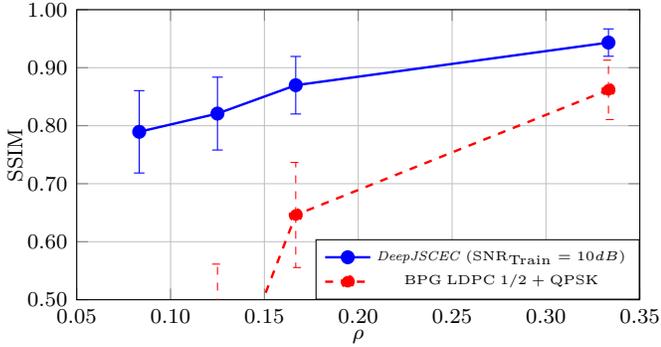

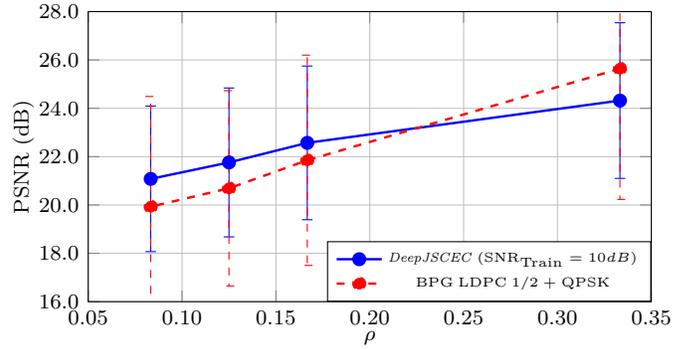
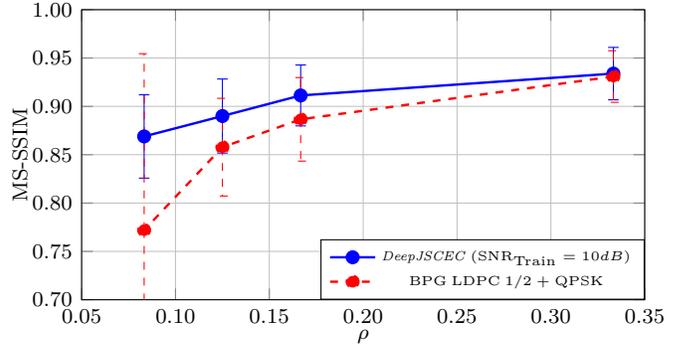
\begin{figure}
    \centering
  \subfloat[PSNR\label{subfig:tinyimagenet_psnr_v_bw}]{%
    \begin{tikzpicture}
        \pgfplotsset{
            legend style={
                font=\fontsize{5}{5}\selectfont,
                at={(1.0,.0)},
                anchor=south east,
            },
            width=0.5\textwidth,
            height=0.3\textwidth,
            xmin=0.05,
            xmax=0.35,
            ymin=16,
            ymax=28,
            xtick distance=0.05,
            ytick distance=2,
            xlabel={$\rho$},
            ylabel={PSNR (dB)},
            grid=both,
            grid style={line width=.1pt, draw=gray!10},
            major grid style={line width=.2pt,draw=gray!50},
            every axis/.append style={
                x label style={
                    font=\fontsize{8}{8}\selectfont,
                    at={(axis description cs:0.5,-0.07)},
                    },
                y label style={
                    font=\fontsize{8}{8}\selectfont,
                    at={(axis description cs:-0.08,0.5)},
                    },
                x tick label style={
                    font=\fontsize{8}{8}\selectfont,
                    /pgf/number format/.cd,
                    fixed,
                    fixed zerofill,
                    precision=2,
                    /tikz/.cd
                    },
                y tick label style={
                    font=\fontsize{8}{8}\selectfont,
                    /pgf/number format/.cd,
                    fixed,
                    fixed zerofill,
                    precision=1,
                    /tikz/.cd
                    },
            }
        }
        \begin{axis}
        \addplot[blue, solid, line width=0.9pt, 
        mark=*, mark options={fill=blue, scale=1.1}, error bars/.cd, y dir=both, y explicit, every nth mark=1] 
        table [x=rho, y=l2, y error=l2_std, col sep=comma]
              {data/djscec_tinyimagenet_nembed16_snr10_rate_dist.csv};
        \addlegendentry{\textit{DeepJSCEC}
        ($\text{SNR}_{\text{Train}}=10dB$)}
        
        \addplot[red, dashed, line width=0.9pt, 
        mark=*, mark options={fill=red, scale=1.1}, error bars/.cd, y dir=both, y explicit, every nth mark=1] 
        table [x=rho, y=l2, y error=l2_std, col sep=comma]
        {data/bpg_tinyimagenet_rate_dist.csv};
        \addlegendentry{BPG LDPC 1/2 + QPSK}
        \end{axis}
        \end{tikzpicture}
    }
    \\
  \subfloat[MS-SSIM\label{subfig:tinyimagenet_msssim_v_bw}]{%
    \begin{tikzpicture}
        \pgfplotsset{
            legend style={
                font=\fontsize{5}{5}\selectfont,
                at={(1.0,0.)},
                anchor=south east,
            },
            width=0.5\textwidth,
            height=0.3\textwidth,
            xmin=0.05,
            xmax=0.35,
            ymin=0.70,
            ymax=1,
            xtick distance=0.05,
            ytick distance=0.05,
            xlabel={$\rho$},
            ylabel={MS-SSIM},
            grid=both,
            grid style={line width=.1pt, draw=gray!10},
            major grid style={line width=.2pt,draw=gray!50},
            every axis/.append style={
                x label style={
                    font=\fontsize{8}{8}\selectfont,
                    at={(axis description cs:0.5,-0.07)},
                    },
                y label style={
                    font=\fontsize{8}{8}\selectfont,
                    at={(axis description cs:-0.08,0.5)},
                    },
                x tick label style={
                    font=\fontsize{8}{8}\selectfont,
                    /pgf/number format/.cd,
                    fixed,
                    fixed zerofill,
                    precision=2,
                    /tikz/.cd
                    },
                y tick label style={
                    font=\fontsize{8}{8}\selectfont,
                    /pgf/number format/.cd,
                    fixed,
                    fixed zerofill,
                    precision=2,
                    /tikz/.cd
                    },
            }
        }
        \begin{axis}
        \addplot[blue, solid, line width=0.9pt, 
        mark=*, mark options={fill=blue, scale=1.1}, error bars/.cd, y dir=both, y explicit, every nth mark=1] 
        table [x=rho, y=msssim, y error=msssim_std, col sep=comma]
        {data/djscec_tinyimagenet_nembed16_snr10_rate_dist.csv};
        \addlegendentry{\textit{DeepJSCEC}
        ($\text{SNR}_{\text{Train}}=10dB$)}
        
        \addplot[red, dashed, line width=0.9pt, 
        mark=*, mark options={fill=red, scale=1.1}, error bars/.cd, y dir=both, y explicit, every nth mark=1] 
        table [x=rho, y=msssim, y error=msssim_std, col sep=comma]
        {data/bpg_tinyimagenet_rate_dist.csv};
        \addlegendentry{BPG LDPC 1/2 + QPSK}
        \end{axis}
        \end{tikzpicture}
        }
  \caption{
  Comparison of \emph{DeepJSCEC} trained on the Tiny ImageNet dataset to BPG compression for different bandwidth compression ratios $\rho$.
  }
  \label{fig:tinyimagenet_distortion_v_bw} 
\end{figure}

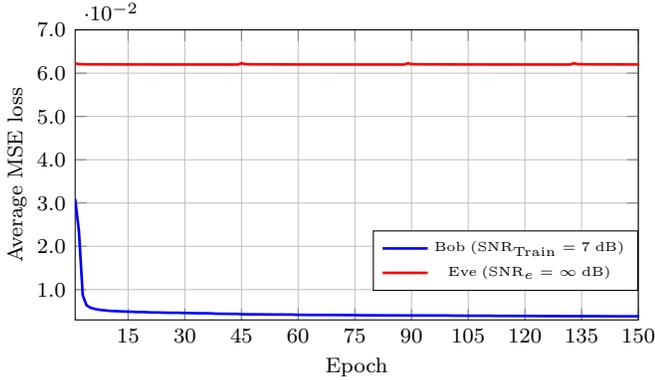
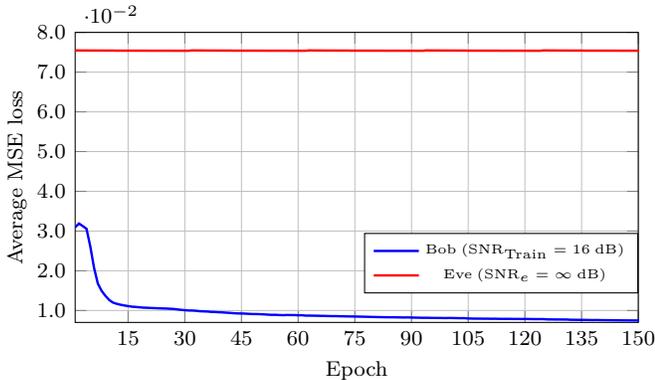
\begin{figure} 
  \centering
  \subfloat[MSE training loss on the CIFAR10 dataset. 
  \label{subfig:cifar10_trainvattack}]{%
    \begin{tikzpicture}
        \pgfplotsset{
            legend style={
                font=\fontsize{5}{5}\selectfont,
                at={(1.0,.1)},
                anchor=south east,
            },
            width=0.5\textwidth,
            height=0.3\textwidth,
            xmin=1,
            xmax=150,
            ymin=0.003,
            ymax=0.07,
            xtick distance=15,
            ytick distance=0.01,
            xlabel={Epoch},
            ylabel={Average MSE loss},
            grid=both,
            grid style={line width=.1pt, draw=gray!10},
            major grid style={line width=.2pt,draw=gray!50},
            every axis/.append style={
                x label style={
                    font=\fontsize{8}{8}\selectfont,
                    at={(axis description cs:0.5,-0.1)},
                    },
                y label style={
                    font=\fontsize{8}{8}\selectfont,
                    at={(axis description cs:-0.07,0.5)},
                    },
                x tick label style={
                    font=\fontsize{8}{8}\selectfont,
                    /pgf/number format/.cd,
                    fixed,
                    fixed zerofill,
                    precision=0,
                    /tikz/.cd
                    },
                y tick label style={
                    font=\fontsize{8}{8}\selectfont,
                    /pgf/number format/.cd,
                    fixed,
                    fixed zerofill,
                    precision=1,
                    /tikz/.cd
                    },
            }
        }
        \begin{axis}
        \addplot[blue, solid, line width=1.0pt, 
        error bars/.cd, y dir=both, y explicit, every nth mark=1] 
        table [x=Epoch, y=train_loss, col sep=comma]
        {data/djscec_cifar10_bw1667_nembed16_snr7_trainvattack.csv};
        \addlegendentry{Bob ($\text{SNR}_{\text{Train}} = 7$ dB)}
        
        \addplot[red, solid, line width=1.0pt, 
        error bars/.cd, y dir=both, y explicit, every nth mark=1] 
        table [x=Epoch, y=attack_loss, col sep=comma]
        {data/djscec_cifar10_bw1667_nembed16_snr7_trainvattack.csv};
        \addlegendentry{Eve ($\text{SNR}_e = \infty$ dB)}
        \end{axis}
        \end{tikzpicture}
    }
    \\
  \subfloat[MSE training loss on the Tiny ImageNet dataset. 
  \label{subfig:tinyimagenet_trainvattack}]{%
    \begin{tikzpicture}
        \pgfplotsset{
            legend style={
                font=\fontsize{5}{5}\selectfont,
                at={(1.0,0.1)},
                anchor=south east,
            },
            width=0.5\textwidth,
            height=0.3\textwidth,
            xmin=1,
            xmax=150,
            ymin=0.007,
            ymax=0.08,
            xtick distance=15,
            ytick distance=0.01,
            xlabel={Epoch},
            ylabel={Average MSE loss},
            grid=both,
            grid style={line width=.1pt, draw=gray!10},
            major grid style={line width=.2pt,draw=gray!50},
            every axis/.append style={
                x label style={
                    font=\fontsize{8}{8}\selectfont,
                    at={(axis description cs:0.5,-0.1)},
                    },
                y label style={
                    font=\fontsize{8}{8}\selectfont,
                    at={(axis description cs:-0.07,0.5)},
                    },
                x tick label style={
                    font=\fontsize{8}{8}\selectfont,
                    /pgf/number format/.cd,
                    fixed,
                    fixed zerofill,
                    precision=0,
                    /tikz/.cd
                    },
                y tick label style={
                    font=\fontsize{8}{8}\selectfont,
                    /pgf/number format/.cd,
                    fixed,
                    fixed zerofill,
                    precision=1,
                    /tikz/.cd
                    },
            }
        }
        \begin{axis}[mark options={solid}]
        \addplot[blue, solid, line width=0.9pt, error bars/.cd, y dir=both, y explicit, every nth mark=1] 
        table [x=Epoch, y=train_loss, col sep=comma]
        {data/djscec_tinyimagenet_bw1667_nembed16_snr16_trainvattack.csv};
        \addlegendentry{Bob ($\text{SNR}_{\text{Train}} = 16$ dB)}
        
        \addplot[red, solid, line width=0.9pt, error bars/.cd, y dir=both, y explicit, every nth mark=1] 
        table [x=Epoch, y=attack_loss, col sep=comma]
        {data/djscec_tinyimagenet_bw1667_nembed16_snr16_trainvattack.csv};
        \addlegendentry{Eve ($\text{SNR}_e = \infty$ dB)}
        \end{axis}
        \end{tikzpicture}
        }
  \caption{
  Comparison of training loss evolution between Bob and Eve $(\rho = 1/6)$.
  Note that Eve's training loss is based on eavesdropping on Alice's transmission after the legitimate transceivers have been trained.
  }
  \label{fig:train_v_attack_loss} 
\end{figure}

\begin{figure} 
  \centering
  \subfloat[CIFAR10
  \label{subfig:cifar10_lossl2_snrtrain10_compare}]{%
  \includegraphics[width=0.9\linewidth]{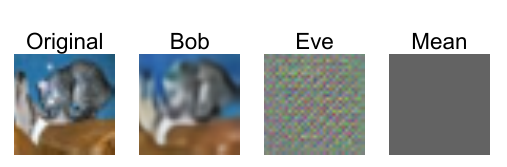}
    }
    \\
  \subfloat[Tiny ImageNet
  \label{subfig:tinyimagenet_lossl2_snrtrain10_compare}]{%
  \includegraphics[width=\linewidth]{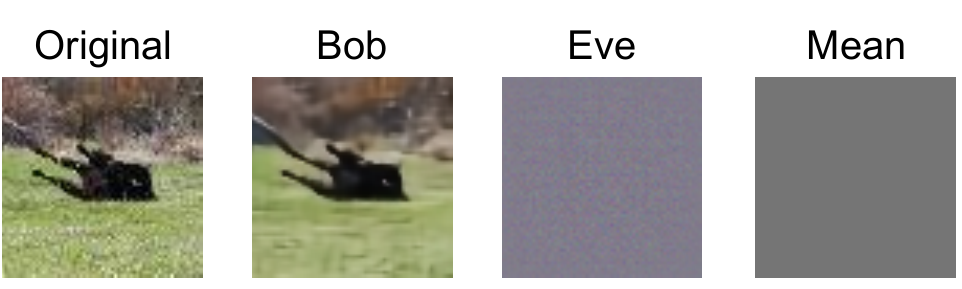}
        }
  \caption{Visual results from models trained on $\text{SNR}_{\text{Train}} = 10$ dB and tested on the same channel SNR using the PSNR metric.}
  \label{fig:visual_comparison} 
  \vspace{-0.3cm}
\end{figure}

\subsection{Wireless Image Transmission}
\label{subsec:reconstruction_results}

For the wireless image transmission problem, described in Sec. \ref{sec:problem_def}, we will consider two datasets, CIFAR10 \cite{CIFARdataset} and Tiny ImageNet \cite{leTinyImageNetVisual}.
For both datasets, we split the samples in the ratio of $0.8:0.1:0.1$ between train, validate and test datasets.
For the CIFAR10 dataset, due to the small image size of $32 \times 32$ relative to Tiny ImageNet's $64 \times 64$, we use a slightly different architecture, where the features are downsampled only 4 times, instead of the 16 times used for Tiny ImageNet.
The network architectures are shown in Table \ref{tab:jscc_image_architectures}.
For the same reason, CIFAR10 was not trained using the MS-SSIM metric.
For both datasets, we use the Adam optimizer \cite{AdamICLR2015} with learning rate $0.0001$ and  parameters $(\beta_1, \beta_2) = (0.9, 0.999)$. 
We also use early stopping with a patience of 10 epochs to prevent overfitting and a learning rate scheduler that multiplies the learning rate by a factor of $0.8$ if the loss does not improve for 5 epochs in a row.

In the architecture, $C$ refers to the number of channels in the output tensor of the convolution operation.
$C_{\text{out}}$ refers to the number of channels in the final output tensor of the encoder $f_{\boldsymbol{\theta}}$, which controls the number of channel uses $k$ per image.
The ``Pixel shuffle" module, within the ``Residual block (RB) upsample" module, is used to increase the height and width of the input tensor by reshaping it, such that the channel dimension is reduced while the height and width dimensions are increased. 
This was first proposed in \cite{shi_real-time_2016} as a less computationally expensive method for increasing the CNN tensor dimensions without requiring a large number of parameters, like transpose convolutional layers.
The GDN layer refers to generalized divisive normalization, initially proposed in \cite{balle2015density}, and shown to be effective in density modeling and compression of images.
The Attention layer refers to the simplified attention module proposed in \cite{cheng_learned_2020}, which reduces the computation cost of the attention module originally proposed in \cite{wang_non-local_2018}.
While the attention layer has been used in \cite{cheng_learned_2020} and \cite{wang_non-local_2018} to improve the compression efficiency by learning to focus on image regions that require higher bit rates, in our model it is used to improve the allocation of channel bandwidth and power resources.

\begin{table*}
\centering
\caption{
Comparison between the image quality obtained for Bob and Eve after training.
Bob's decoder was trained jointly with the encoder for $\text{SNR}_{\text{Train}} = 10$ dB and evaluated at the same SNR.
The mean results are based on averaging the test dataset pixels and corresponds to the image quality if the pixel values are all equal to the average pixel value.
}
\label{tab:bob_vs_eve_reconstruction}
\begin{tabular}{|c|cc|cc|cc|}
\hline
\multirow{2}{*}{datasets} & \multicolumn{2}{c|}{Bob}    & \multicolumn{2}{c|}{Eve}    & \multicolumn{2}{c|}{$\hat{\mathbf{x}} = \mathbb{E}[\mathbf{x}]$}    \\ \cline{2-7} 
                  & \multicolumn{1}{c|}{PSNR} & (MS-)SSIM & \multicolumn{1}{c|}{PSNR} & (MS-)SSIM & \multicolumn{1}{c|}{PSNR} & (MS-)SSIM \\ \hline
                CIFAR10  & \multicolumn{1}{c|}{$25.95$ dB} & $0.8698$ & \multicolumn{1}{c|}{$12.29$ dB} & $0.0624$ & \multicolumn{1}{c|}{$12.42$ dB} & $0.0860$ \\ \hline
                TinyImageNet  & \multicolumn{1}{c|}{$22.57$ dB} & $0.9113$ & \multicolumn{1}{c|}{$11.28$ dB} & $0.1803$ & \multicolumn{1}{c|}{$12.73$ dB} & $0.1965$ \\ \hline
\end{tabular}
\end{table*}

We will consider models trained for different $\text{SNR}_{\text{Train}}$ and tested over a range of channel SNRs.
That is, during training, the channel SNR is fixed to $\text{SNR}_{\text{Train}}$, while during testing, the channel SNR varies.
The $\text{SNR}_{\text{Train}}$ values were chosen such that it coincides with the SNR where the LDPC codes and modulation orders we consider herein would fail, in order to highlight the behavior of \emph{DeepJSCEC} around the cliff-edge of the digital baseline.
Fig. \ref{fig:cifar10_reconstruction_graceful} shows the performance of the models trained on the CIFAR10 dataset and tested over a range of channel SNRs for Bob.
It can be seen that the performance of \emph{DeepJSCEC} gracefully degrades as the test SNR decreases, which is in contrast to the cliff-edge deterioration of the digital schemes at their respective threshold SNRs. 
This is true when we optimize for either the PSNR or the SSIM metric.
Moreover, the image quality from \emph{DeepJSCEC} for a given SNR is much higher than the digital schemes and remains superior across a range of bandwidth compression ratios, as shown in Fig. \ref{fig:cifar10_distortion_v_bw}, demonstrating the theoretical superiority of JSCC \cite{Gastpar:IT:03, kostina_lossy_2013, gallager_information_1968}.
We note that although these characteristics have been shown previously by \cite{Eirina:TCCN:19, Kurka:IZS2020, Kurka:deepjsccf:jsait, Kurka:BandwidthAgile:TWComm2021, tungDeepWiVeDeepLearningAidedWireless2021}, this is the first time an end-to-end encrypted JSCC scheme, implemented through DNNs, has shown these properties.

Similar results are shown in Fig. \ref{fig:tinyimagenet_reconstruction_graceful} and \ref{fig:tinyimagenet_distortion_v_bw} for models trained on the Tiny ImageNet dataset, where we consider the PSNR and MS-SSIM metrics, as the latter is more perceptually aligned to human vision and the image size of the dataset supports it.
The only exception is in the high SNR and high bandwidth compression ratio regime, where \emph{DeepJSCEC} fell below the digital baseline for $\text{SNR}_b > 16$ dB  or $\rho = 1/3$ at $\text{SNR}_b = 10$ dB when optimized for the PSNR metric.
In all other scenarios, \emph{DeepJSCEC} is superior to the digital baseline, and is better when optimized and evaluated on the more perceptually aligned MS-SSIM metric.

To demonstrate the security properties of \emph{DeepJSCEC}, we devise a chosen-plaintext attack, where Eve has access to the encoder function $f_{\boldsymbol{\theta}}$, the public key $\mathbf{P}(\mathbf{S})$, and the dataset of images used to train $(f_{\boldsymbol{\theta}}, g_{\boldsymbol{\phi}})$.
This allows Eve to obtain pairs of $(\mathbf{x}, \mathbf{c})$ and the objective for Eve is to use these pairs to obtain the secret key $\mathbf{S}$ so that any image transmitted using $f_{\boldsymbol{\theta}}$ and the encryption scheme $E$ using the public key $\mathbf{P}(\mathbf{S})$ can be deciphered.
Due to the complexity of search algorithms for finding $\mathbf{S}$ based on $(\mathbf{x}, \mathbf{c})$, we will instead train a DNN decoder $\bar{\mathbf{x}} = \bar{g}_{\boldsymbol{\psi}}(\mathbf{c})$, parameterized by $\boldsymbol{\psi}$, and attempt to optimize the parameters $\boldsymbol{\psi}$ so that $d(\mathbf{x}, \bar{\mathbf{x}})$, as measured by either the PSNR, SSIM or MS-SSIM metric, is maximized. 
This is akin to the cryptographic game defined in Def. \ref{def:ind_cpa}, where we essentially use a DNN to parameterize the secret key and search with gradient descent.
We note that this corresponds to infinite channel SNR ($\text{SNR}_e = \infty$) for Eve, as the ciphertexts are obtained without any additional noise from the channel.
This means the results of this chosen-plaintext attack acts as an upper bound on how well Eve is expected to decode the intercepted signal.

\begin{table*}
\centering
\caption{
Network architectures used by \emph{DeepJSCEC} and the feature compression network for the remote classification problem.
The block definitions are shown in Fig. \ref{fig:residual_blocks}.
}
\label{tab:jscc_remote_classification}
\begin{tabular}{|cc|cc|}
\hline
\multicolumn{2}{|c|}{\emph{DeepJSCEC}}    & \multicolumn{2}{c|}{Feature Compression Network}    \\ \hline
\multicolumn{1}{|c|}{$f_{\boldsymbol{\theta}}$} & $g_{\boldsymbol{\phi}}$ & \multicolumn{1}{c|}{$f_{\boldsymbol{\theta}^\prime}$} & $g_{\boldsymbol{\phi^\prime}}$ \\ \hline
\multicolumn{1}{|c|}{RB stride 2 (256)} & RB (256) & \multicolumn{1}{c|}{RB stride 2 (256)} & RB (256) \\ \hline
\multicolumn{1}{|c|}{RB (256)} & RB (512) & \multicolumn{1}{c|}{RB (256)} & RB (512) \\ \hline
\multicolumn{1}{|c|}{Attention block (256)} & Attention block (512) & \multicolumn{1}{c|}{Attention block (256)} & Attention block (512) \\ \hline
\multicolumn{1}{|c|}{RB stride 2 (256)} & RB (512) & \multicolumn{1}{c|}{RB stride 2 (256)} & RB (512) \\ \hline
\multicolumn{1}{|c|}{RB (256)} & RB stride 2 (1024) & \multicolumn{1}{c|}{RB (256)} & RB stride 2 (1024) \\ \hline
\multicolumn{1}{|c|}{Attention block ($C_{\text{out}}$)} & RB (2048) & \multicolumn{1}{c|}{Attention block (256)} & RB (2048) \\ \hline
\multicolumn{1}{|c|}{} & RB (2048) & \multicolumn{1}{c|}{} & RB (2048) \\ \hline
\multicolumn{1}{|c|}{} & Attention block (2048) & \multicolumn{1}{c|}{} & Attention block (2048) \\ \hline
\multicolumn{1}{|c|}{} & Average Pooling 2D ($4 \times 4$) & \multicolumn{1}{c|}{} & Average Pooling 2D ($4 \times 4$) \\ \hline
\multicolumn{1}{|c|}{} & Linear (2048) & \multicolumn{1}{|c|}{} & Linear (2048)\\ \hline
\multicolumn{1}{|c|}{} & Leaky ReLU & \multicolumn{1}{|c|}{} & Leaky ReLU\\ \hline
\multicolumn{1}{|c|}{} & Linear (10) & \multicolumn{1}{|c|}{} & Linear (10)\\ \hline
\multicolumn{1}{|c|}{} & Softmax & \multicolumn{1}{|c|}{} & Softmax \\ \hline
\end{tabular}
\end{table*}

Fig. \ref{fig:train_v_attack_loss} shows the MSE training loss for the legitimate receiver Bob versus the eavesdropper Eve.
As Bob's average MSE loss decreases over training epochs and eventually converges to the results we see in Figs. \ref{fig:cifar10_reconstruction_graceful} and \ref{fig:tinyimagenet_reconstruction_graceful}, Eve's training loss never decreases, showing that the encryption scheme is indeed working as intended.
We show in Table \ref{tab:bob_vs_eve_reconstruction}, the resultant image quality for Bob, Eve and a baseline that generates images using the average pixel values of the test dataset. 
We see that while Bob is able to reconstruct the transmitted image with good quality, Eve achieves a result slightly worse than reconstructing an image with average pixel values.
This is a strong indication that Eve is not able to learn a decoder that performs significantly better than a random sample from the source distribution.
Fig. \ref{fig:visual_comparison} shows this observation visually and confirms that Eve's reconstruction bears no similarity with the original image.

\subsection{Remote Classification}
\label{subsec:classification_results}

In this section, we will show that the proposed joint source-channel and encryption coding method can also be used for a remote classification problem.
The idea is that an edge device with insufficient computational resources wants to offload a classification task to a remote server that can classify the image and transmit the label back.
Let us define the edge device as Alice and the remote server as Bob.
In the context of this paper, Eve is an eavesdropper that wants to intercept the messages transmitted by Alice in order to obtain the image label that Alice wants to classify.
We will use the CIFAR10 dataset as before to demonstrate this, and modify the encoder and decoder architectures to accommodate for this task.
The architectures used herein are shown in Table \ref{tab:jscc_remote_classification}, and share the same blocks as the ones used in the image transmission problem in Sec. \ref{subsec:reconstruction_results}, except with a bias towards larger decoder networks than the encoder.
This accounts for the lower computational resources of the edge device than the remote server.
We will again assume that Eve has the same amount of computational resources as Bob with the same network architecture.

Since the objective is to classify the images, we use the cross entropy (CE) loss, instead of the reconstruction losses in Sec. \ref{sec:problem_def}, defined as
\begin{equation}
    l_{\text{CE}} = -\sum_{j=1}^{N_C} \mathbb{I}_{(T(\mathbf{x}) = j)} g_{\boldsymbol{\phi}}(\hat{\mathbf{z}})_j,
\end{equation}
where $N_C$ is the number of classes, $T(\mathbf{x})$ is the integer class label for the image $\mathbf{x}$, and $\mathbb{I}_{(T(\mathbf{x}) = j)}$ is the indicator function for the event $T(\mathbf{x}) = j$, which corresponds to an one-hot encoding of the image label.

In order to compare to a digital baseline, we will train a lossy feature compression network to learn to transmit only the essential features for classification within a given bit rate.
That is, we train a DNN for classifying the CIFAR10 dataset with an objective function that trades off accuracy and the number of bits needed to represent the features at the output of the DNN encoder.
The network architecture we use for this task is described in Table \ref{tab:jscc_remote_classification} under ``Feature Compression Network".

In order to estimate the number of bits needed to represent the output features, we quantize the encoder output by rounding to the nearest integer and use a flexible Gaussian mixture model to estimate the probability mass function (PMF) of the quantized output.
Then, to enable an end-to-end differentiable approach, we utilize the well-known quantization noise \cite{grayQuantization2006a} to model the quantization process.
Specifically, let $\mathbf{z} = f_{\boldsymbol{\theta}^\prime}(\mathbf{x})$, then in the training phase we add the uniform noise to each element of the latent representation instead of rounding to the nearest integer, as follows:
\begin{equation}
    q(\mathbf{z}) = \mathbf{z} + \mathcal{U}(-0.5, 0.5),
\end{equation}
where $q(\cdot)$ is the approximate quantization operation and $\mathcal{U}(a, b)$ is the uniform distribution with support in $[a, b]$.
This formulation gives a good approximation of the rounding operation performed during inference.
Note that this is only used for the digital baseline and is different from the quantizer $q_{\mathcal{C}_q}$ used in \emph{DeepJSCEC}.
The distribution of the features are then estimated using a simple yet flexible Gaussian mixture model to estimate the PMF. 
That is, 
we estimate the discrete probability of the $i$th value in $\Tilde{\mathbf{z}} = q(\mathbf{z})$ during training as
\begin{align}
    &P(\Tilde{z}_i) \nonumber \\
    &= \sum_{j=1}^K \alpha_j \left( Q \left( \frac{(\Tilde{z}_i - 0.5) - \mu_j}{\sigma_j} \right) \nonumber - Q \left( \frac{(\Tilde{z}_i + 0.5) - \mu_j}{\sigma_j} \right) \right),
\end{align}
where $Q(x) = \int_{x}^\infty \frac{1}{\sqrt{2\pi}} e^{\frac{-x^2}{2}}$ is the standard Gaussian tail distribution function and $K$ is the number of mixtures.
During inference, we use the rounded values $\bar{\mathbf{z}} = \lfloor \mathbf{z} \rceil$ to compute the probabilities $P(\bar{\mathbf{z}})$.
Having this distribution allows us to use an entropy code, such as arithmetic coding \cite{mackayInformationTheoryInference2003} to compress the features to the rate estimated by the mixture model.
The training loss for the feature compression model is then described by
\begin{equation}
    L = l_{\text{CE}} - \lambda \log_2 \left( P(\Tilde{\mathbf{z}}) \right),
\end{equation}
where $\lambda$ is a Lagrangian multiplier that trades off prediction accuracy with rate of compression.
In this experiment, we use $K=9$ mixtures.

We remark that, here we learn the distribution of the quantized feature vectors $\bar{\mathbf{z}}$, but unlike recent works \cite{balleNonlinearTransformCoding2021}, we do not consider adaptive probability models or use another neural network to predict parameters $\{(\alpha_j, \mu_j, \sigma_j) \}_{j=1}^K$ of the mixture. 
This is because the proposed simple model performs sufficiently well, and has been used in prior works that consider similar problems \cite{jankowskiWirelessImageRetrieval2021}. 

\begin{table}
\centering
\caption{
Comparison between the test accuracy obtained for Bob and Eve after training.
Bob's decoder was trained jointly with the encoder for $\text{SNR}_{\text{Train}} = 10$ dB and evaluated at the same SNR.
The random results are based on randomly guessing the image class uniformly.
}
\label{tab:bob_vs_eve_classification}
\begin{tabular}{|c|c|c|}
\hline
Bob & Eve & Random \\ \hline
$0.7507$ & $0.0938$ & $0.0978$ \\ \hline
\end{tabular}
\end{table}

\begin{figure} 
\centering
\begin{tikzpicture}
    \pgfplotsset{
        legend style={
            font=\fontsize{5}{5}\selectfont,
            at={(1.0,0.)},
            anchor=south east,
        },
        height=0.6\textwidth,
        width=0.5\textwidth,
        xmin=2,
        xmax=17,
        ymin=0.6,
        ymax=0.78,
        xtick distance=2,
        ytick distance=0.02,
        xlabel={$\text{SNR}_b$ (dB)},
        ylabel={Test accuracy},
        grid=both,
        grid style={line width=.1pt, draw=gray!10},
        major grid style={line width=.2pt,draw=gray!50},
        every axis/.append style={
            x label style={
                font=\fontsize{8}{8}\selectfont,
                at={(axis description cs:0.5, -0.04)},
                },
            y label style={
                font=\fontsize{8}{8}\selectfont,
                at={(axis description cs:-0.08,0.5)},
                },
            x tick label style={
                font=\fontsize{8}{8}\selectfont,
                /pgf/number format/.cd,
                fixed,
                fixed zerofill,
                precision=0,
                /tikz/.cd
                },
            y tick label style={
                font=\fontsize{8}{8}\selectfont,
                /pgf/number format/.cd,
                fixed,
                fixed zerofill,
                precision=2,
                /tikz/.cd
                },
        }
    }
    \begin{axis}
    \addplot[green, solid, line width=0.9pt, mark=*, mark options={fill=green, scale=1.1}, error bars/.cd, y dir=both, y explicit, every nth mark=1] 
    table [x=SNR, y=snr_train7, col sep=comma]
    {data/djscec_cifar10_bw1667_nembed16_acc.csv};
    \addlegendentry{\textit{DeepJSCEC} 
    ($\text{SNR}_{\text{Train}}=7dB$)}
    
    \addplot[red, solid, line width=0.9pt, mark=triangle*, mark options={fill=red, scale=1.1}, error bars/.cd, y dir=both, y explicit, every nth mark=1] 
    table [x=SNR, y=snr_train10, col sep=comma]
    {data/djscec_cifar10_bw1667_nembed16_acc.csv};
    \addlegendentry{\textit{DeepJSCEC} 
    ($\text{SNR}_{\text{Train}}=10dB$)}
    
    \addplot[blue, solid, line width=0.9pt, mark=square*, mark 
    options={fill=blue, scale=1.1, solid}, error bars/.cd, y dir=both, y explicit, every nth mark=1] 
    table [x=SNR, y=snr_train12, col sep=comma]
    {data/djscec_cifar10_bw1667_nembed16_acc.csv};
    \addlegendentry{\textit{DeepJSCEC} 
    ($\text{SNR}_{\text{Train}}=12dB$)}
    
    \addplot[teal, solid, line width=0.9pt, mark=square*, mark 
    options={fill=teal, scale=1.1, solid}, error bars/.cd, y dir=both, y explicit, every nth mark=1] 
    table [x=SNR, y=snr_train16, col sep=comma]
    {data/djscec_cifar10_bw1667_nembed16_acc.csv};
    \addlegendentry{\textit{DeepJSCEC} 
    ($\text{SNR}_{\text{Train}}=16dB$)}
    
    \addplot[color=black, dashed, line width=1.2pt, mark=*, mark options={fill=black, solid, scale=1.1}, 
    error bars/.cd, y dir=both, y explicit, every nth mark=1] 
    table [x=SNR, y=1_2_bpsk, col sep=comma]
    {data/digital_feature_compress_cifar10_bw1667.csv};
    \addlegendentry{Digital + LDPC 1/2 BPSK $( \lambda = 4 \times 10^{-4} )$}
    
    \addplot[color=magenta, dashed, line width=1.2pt, 
    mark=*, mark options={fill=magenta, solid, scale=1.1}, 
    error bars/.cd, y dir=both, y explicit, every nth mark=1] 
    table [x=SNR, y=3_4_bpsk, col sep=comma]
    {data/digital_feature_compress_cifar10_bw1667.csv};
    \addlegendentry{Digital + LDPC 3/4 BPSK $( \lambda = 2 \times 10^{-4} )$}
    
    \addplot[color=black, dashed, line width=1.2pt, 
    mark=triangle*, mark options={fill=black, solid, scale=1.1}, 
    error bars/.cd, y dir=both, y explicit, every nth mark=1] 
    table [x=SNR, y=1_2_qpsk, col sep=comma]
    {data/digital_feature_compress_cifar10_bw1667.csv};
    \addlegendentry{Digital + LDPC 1/2 QPSK $( \lambda = 10^{-4} )$}
    
    \addplot[color=magenta, dashed, line width=1.2pt, 
    mark=triangle*, mark options={fill=magenta, solid, scale=1.1}, 
    error bars/.cd, y dir=both, y explicit, every nth mark=1] 
    table [x=SNR, y=3_4_qpsk, col sep=comma]
    {data/digital_feature_compress_cifar10_bw1667.csv};
    \addlegendentry{Digital + LDPC 3/4 QPSK $( \lambda = 5 \times 10^{-5} )$}
    
    \addplot[color=black, dashed, line width=1.2pt, 
    mark=square*, mark options={fill=black, solid, scale=1.1}, 
    error bars/.cd, y dir=both, y explicit, every nth mark=1] 
    table [x=SNR, y=1_2_16qam, col sep=comma]
    {data/digital_feature_compress_cifar10_bw1667.csv};
    \addlegendentry{Digital + LDPC 1/2 16QAM $( \lambda = 3 \times 10^{-5} )$}
    \end{axis}
\end{tikzpicture}
    
\caption{
Comparison between \emph{DeepJSCEC} trained on the CIFAR10 dataset for different $\text{SNR}_{\text{Train}}$ and digital baselines using learned feature compression, AES for encryption and LDPC codes for channel coding, for the remote classification problem ($\rho = 1/6$).
}
\label{fig:cifar10_categorical_graceful} 
\end{figure}
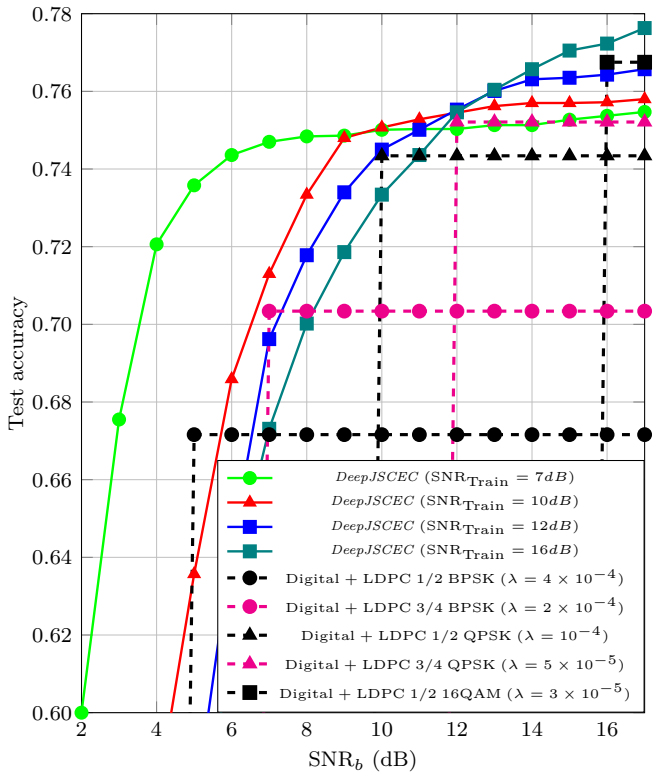

Fig. \ref{fig:cifar10_categorical_graceful} shows the comparison between \emph{DeepJSCEC} and the digital baseline using the learned digital compression network, for the remote classification task.
We see that a similar trend to wireless image transmission problem in the previous section, can be seen here.
The classification accuracy smoothly adjusts to the channel SNR, while the digital baseline exhibits the cliff-effect due to the thresholding effects of the channel code.
\emph{DeepJSCEC} also achieves higher classification accuracy than the digital baseline for all SNRs considered.
Moreover, when we implemented the same chosen-plaintext attack on this problem, we observe that Eve is only able to obtain a classification accuracy close to a random guess, as shown in Table \ref{tab:bob_vs_eve_classification}.
This is further evidence that the encryption scheme used in \emph{DeepJSCEC} is IND-CPA secure and that it is problem agnostic, making our proposed solution readily extendable to other problem domains.

\section{Conclusions}
\label{sec:conclusions}

In this paper we have proposed \emph{DeepJSCEC}, the first DNN-driven JSCC scheme for wireless image transmission that is secure against an eavesdropper without making any assumptions on its channel quality or the intended use of the intercepted signal.
\emph{DeepJSCEC} not only achieves better end-to-end distortion than the conventional separation-based digital schemes, using BPG for compression, AES for encryption, and LDPC codes for channel coding, but also provides graceful degradation of image quality in varying channel conditions.
The security of \emph{DeepJSCEC} is measured under the IND-CPA criterion and it is shown that the proposed solution provides an advantage close to 0 to the eavesdropper, without any assumptions on the eavesdropper's channel condition.
Moreover, we also show that the solution can also be applied to other semantic communication problems, such as remote classification at the edge, where \emph{DeepJSCEC} also shows superior classification accuracy than the digital baseline. 
The proposed solution is highly practical and readily extendable to other end-to-end JSCC problems.

\bibliographystyle{ieeetr}
\bibliography{references.bib}

\end{document}